\documentclass[aps,pra,showpacs,floatfix,superscriptaddress]{revtex4}
\usepackage{graphicx}
\usepackage{subfigure}
\usepackage{color}
\usepackage{ulem}

\begin{document}

\title{Dynamics of entanglement in Two-Qubit Open System Interacting with a
Squeezed Thermal Bath via Dissipative interaction}

\author{Subhashish Banerjee}
\email{subhashish@cmi.ac.in}
\affiliation{Raman Research Institute, Bangalore- 560080, India}
\affiliation{Chennai Mathematical Institute, Padur PO, Siruseri- 603103, India}
\author{V. Ravishankar} 
\email{vravi@iitk.ac.in}
\affiliation{Raman Research Institute, Bangalore- 560080, India}
\affiliation{Indian Institute of Technology, Kanpur, India} 
\author{R. Srikanth}
\email{srik@rri.res.in}
\affiliation{Poornaprajna Institute for Scientific Research, 
Bangalore- 560080, India}
\affiliation{Raman Research Institute, Bangalore- 560080, India}


\begin{abstract}
  We  study  the  dynamics  of  entanglement  in  a  two-qubit  system
  interacting  with   a  squeezed  thermal  bath   via  a  dissipative
  system-reservoir interaction  with the system  and reservoir assumed
  to be in  a separable initial state.  The  resulting entanglement is
  studied  by making  use of  a recently  introduced measure  of mixed
  state entanglement via a  probability density function which gives a
  statistical  and  geometrical  characterization of  entanglement  by
  exploring the entanglement content in the various subspaces spanning
  the  two-qubit Hilbert  space. We  also make  an application  of the
  two-qubit  dissipative dynamics  to  a simplified  model of  quantum
  repeaters.
\end{abstract} 


\maketitle

\section{Introduction}

Open quantum systems  take into account the effect  of the environment
(reservoir  or bath)  on  the  dynamical evolution  of  the system  of
interest thereby providing a  natural route for discussing damping and
dephasing. One of the first  testing grounds for open system ideas was
in quantum  optics \cite{wl73}. Its application to  other areas gained
momentum from the works of Caldeira and Leggett \cite{cl83}, and Zurek
\cite{wz93}, among others.  The total Hamiltonian  is $H = H_S + H_R +
H_{SR}$, where  $S$ stands for the system, $R$  for the reservoir and
$SR$  for  the   system-reservoir  interaction.   Depending  upon  the
system-reservoir  ($S-R$)  interaction, open  systems  can be  broadly
classified into two categories,  viz., quantum non-demolition (QND) or
dissipative.  A  particular type of quantum  nondemolition (QND) $S-R$
interaction is  given by a class of  energy-preserving measurements in
which dephasing occurs without  damping the system, i.e., where $[H_S,
H_{SR}]  = 0$  while the  dissipative systems  correspond to  the case
where  $[H_S, H_{SR}]  \neq  0$ resulting  in  decoherence along  with
dissipation \cite{bg07}.

A  prototype   of  dissipative  open  quantum   systems,  having  many
applications, is the quantum  Brownian motion of harmonic oscillators.
This model  was studied  by Caldeira and  Leggett \cite{cl83}  for the
case where  the system and  its environment were  initially separable.
The above treatment of the  quantum Brownian motion was generalized to
the physically  reasonable initial condition  of a mixed state  of the
system and its environment  by Hakim and Ambegaokar \cite{ha85}, Smith
and  Caldeira \cite{sc87}, Grabert,  Schramm and  Ingold \cite{gsi88},
and for the case of a system in a Stern-Gerlach potential \cite{sb00},
and   also   for   the   quantum  Brownian   motion   with   nonlinear
system-environment couplings  \cite{sb03-2}, among others.  

The  interest  in  the  relevance  of open  system  ideas  to  quantum
information has  increased in recent  times because of  the impressive
progress made on the experimental front in the manipulation of quantum
states of  matter towards  quantum information processing  and quantum
communication.  Myatt {\it et  al.} \cite{myatt} and Turchette {\it et
al.} \cite{turch} have performed a series of experiments in which they
induced decoherence and decay  by coupling the atom (their system-$S$)
to engineered reservoirs, in which  the coupling to, and the state of,
the  environment  are controllable. 

Quantum entanglement is  the inherent property of a  system to exhibit
correlations, the physical basis being the non-local nature of quantum
mechanics  \cite{bell}, and hence  is a  property that  is exclusively
quantum  in  nature. Entanglement  plays  a  central  role in  quantum
information   theory  \cite{nc}   as   in  interesting   non-classical
applications such as quantum computation \cite{shor} and quantum error
correction  \cite{css}. A  number of  methods have  been  proposed for
creating  entanglement  involving  trapped  atoms  \cite{beige,  fy00,
sm02}.

An important issue is to study how quantum entanglement is affected by
noise, which  can be thought of  as a manifestation of  an open system
effect  \cite{bp02}.   In   \cite{wlk8}  entanglement  of  a  two-mode
squeezed state  in a phase-sensitive Gaussian  environment was studied
and  the  criteria for  the  necessary  and  sufficient condition  for
separability of  Gaussian continuous-variable states  \cite{simon} was
employed as a measure of entanglement. In \cite{pa08} the entanglement
between charge qubits induced  by a common dissipative environment was
analyzed using  concurrence as the measure.   Some recent experimental
investigations  on the  influence of  decoherence on  the  dynamics of
entanglement have been  made in \cite{am07, lc07}.  In  a related work
\cite{brs1}, this  issue was taken up  with the noise  coming from the
effect of the  environment modelled by a QND  $S-R$ interaction.  Here
we complement this  program by studying the effect  of noise, modelled
by a  dissipative $S-R$ interaction  with the reservoir in  an initial
squeezed-thermal   state   \cite{bg07,sqgen},   on  the   entanglement
evolution between two spatially separated (and initially uncorrelated)
qubits, brought  out by interaction with  the bath.  This  would be of
relevance to evaluate the  performance of two-qubit gates in practical
quantum information processing systems.

Since we are  dealing here with a two qubit  system which very rapidly
evolves into a mixed state,  a study of entanglement would necessarily
involve a measure of entanglement for mixed states.  Entanglement of a
bipartite  system in  a pure  state is  unambigious and  well defined.
However, mixed state entanglement (MSE)  is not so well defined. Thus,
although  a  number of  criteria  such  as  entanglement of  formation
\cite{bd96, ww98, mc05} and separability \cite{fw89} exist, there is a
realization  \cite {bd96}  that  a single  quantity  is inadequate  to
describe MSE. This was the principal motivation for the development of
a new prescription of MSE \cite{br08} in which it is characterized not
as a function, but as a probability density function (PDF).  The known
prescriptions such as concurrence  and negativity emerge as particular
parameters  that  characterize   the  probability  density.   We  will
principally make use  of this measure in our  study of entanglement in
the two-qubit system.

The plan  of the paper is  as follows. In Section  II, we recapitulate
for consistency,  the recently  developed entanglement measure  of MSE
\cite{br08}.   In  Section III,  the  master  equation describing  the
dynamical  evolution  of  the  two-qubit  system  interacting  with  a
squeezed thermal  bath, via a dissipative $S-R$  interaction, is given
which is then  used in Section IV, to study in  detail the dynamics of
the system interacting with a  vacuum bath with zero bath squeezing in
Section  IV(A) and  with a  general squeezed  thermal bath  in Section
IV(B). Section V deals with the entanglement analysis of the two-qubit
open system using the PDF as  a measure of entanglement. We compare it
with the usual measure of MSE, concurrence.  We dwell on the scenarios
where  the  two  qubits   effectively  interact  via  localized  $S-R$
interactions, called  the independent decoherence model,  as also when
they  interact  collectively  with  the bath,  called  the  collective
decoherence model.  In Section VI,  we make a brief application of the
model to  practical quantum  communication, in the  form of  a quantum
repeater   \cite{chb96,bdcz98}.    In  Section   VII,   we  make   our
conclusions.

\section{Characterization of Mixed State Entanglement through a 
Probability Density Function}

Here  we  briefly recapitulate  the  characterization  of mixed  state
entanglement  (MSE) through  a PDF  as developed  in  \cite{br08}.  As
pointed out in the Introduction,  the above criterion was evolved from
the  motivation  that  for  the  characterization  of  MSE,  a  single
parameter  is inadequate.  The  basic idea  is to  express the  PDF of
entanglement  of  a given  system  density  matrix  (in this  case,  a
two-qubit) in  terms of  a weighted sum  over the PDF's  of projection
operators  spanning  the full  Hilbert  space  of  the system  density
matrix.  The PDF of a system in a state which is a projection operator
$\rho = \frac{1}{M}\Pi_M$ of rank $M$ is defined as:
\begin{equation}
{\cal P}_{\Pi_M}({\cal E}) = \frac{\int d{\cal H}_{\Pi_M}\delta
({\cal E}_{\psi} - {\cal E})}{\int d{\cal H}_{\Pi_M}}, \label{entproject}
\end{equation}
where  $\int  d{\cal H}_{\Pi_M}$  is  the  volume  measure for  ${\cal
H}_{\Pi_M}$,  which is the  subspace spanned  by $\Pi_M$.   The volume
measure is  determined by the  invariant Haar measure  associated with
the  group of automorphisms  of $\int  d{\cal H}_{\Pi_M}$,  modulo the
stabilizer group of the reference state generating ${\cal H}_{\Pi_M}$.
Thus for  a one dimensional  projection operator, representing  a pure
state,  the  group of  automorphisms  consists  of  only the  identity
element and  the PDF is  simply given by  the Dirac delta.  Indeed, if
$\rho= \Pi_1  \equiv \vert \psi  \rangle \langle \psi \vert$,  the PDF
has the  form ${\cal  P}_{\rho}({\cal E}) =  \delta ({\cal E}  - {\cal
E}_{\psi})$
thereby resulting  in the description  of pure state  entanglement, as
expected, by a single number.  The entanglement density of a system in
a general  mixed state  $\rho$ is  given by resolving  it in  terms of
nested projection operators with appropriate weights as
\begin{eqnarray} 
\rho &=& (\lambda_1 - \lambda_2)\Pi_1 + (\lambda_2 - \lambda_3)\Pi_2
+ .......(\lambda_{N-1} - \lambda_N)\Pi_{N-1} + \lambda_N \Pi_N  \nonumber\\
&\equiv& \sum_{M=1}^N \Lambda_M \Pi_M, \label{nested}
\end{eqnarray}
where  the projections  are  $\Pi_M =\sum_{j=1}^M|\psi_j\rangle\langle
\psi_j|$,  with $M  = 1,2,...,N$  and the  eigenvalues  $\lambda_1 \ge
\lambda_2  \ge  ....$,  i.e.,   the  eigenvalues  are  arranged  in  a
non-increasing fashion. Thus the PDF for the entanglement of $\rho$ is
given by
\begin{equation}
{\cal P}_{\rho} ({\cal E}) =  \sum_{M=1}^N \omega_M {\cal P}_{\Pi_M}
({\cal E}), \label{PDF}
\end{equation}
where  the weights  of  the respective  projections ${\cal  P}_{\Pi_M}
({\cal E})$ are given by  $\omega_M = \Lambda_M/\lambda_1$.  For a two
qubit system, the density matrix  would be represented as a nested sum
over  four projection  operators, $\Pi_1$,  $\Pi_2$,  $\Pi_3$, $\Pi_4$
corresponding  to one,  two, three  and four  dimensional projections,
respectively, with  $\Pi_1$ corresponding to a pure  state and $\Pi_4$
corresponding  to a  a uniformly  mixed state,  is a  multiple  of the
identity  operator.   The most  interesting  structure  is present  in
$\Pi_2$,  the two-dimensional  projection, which  is  characterized by
three parameters,  viz. ${\cal  E}_{cusp}$, the entanglement  at which
the PDF  diverges, ${\cal E}_{max}$, the  maximum entanglement allowed
and  ${\cal P}_2 ({\cal  E}_{max})$, the  PDF corresponding  to ${\cal
  E}_{max}$. The three dimensional projection $\Pi_3$ is characterized
by the parameter ${\cal E}_{\perp}$, which parametrizes a discontunity
in  the  entanglement  density  function  curve.   By  virtue  of  the
convexity of  the sum over  the nested projections  (\ref{nested}), it
can be seen  that the concurrence of any state $\rho$  is given by the
inequality   ${\cal  C}_{\rho}   \leq  (\lambda_1   -  \lambda_2){\cal
  C}_{\Pi_1} +  (\lambda_2 - \lambda_3){\cal  C}_{\Pi_2}$.  Thus while
the  concurrence  for  a  three  and four  dimensional  projection  is
identically  zero, through the  PDF one  is able  to make  a statement
about the entanglement content of these spaces.  The fact that the PDF
(\ref{PDF}) enables  us to study  entanglement of a physical  state by
exploiting  the richness  inherent  in the  subspaces  spanned by  the
system Hilbert space makes  it an attractive statistical and geometric
characterization of entanglement, of which an explicit illustration is
made in Section V.

\section{Two-Qubit Dissipative interaction with a Squeezed 
Thermal Bath}

We consider the Hamiltonian, describing the dissipative interaction of
$N$ qubits (two-level atomic system)  with the bath (modelled as a 3-D
electromagnetic field (EMF)) via the dipole interaction as \cite{ft02}
\begin{eqnarray}
H & = & H_S + H_R + H_{SR} \nonumber \\ & = & \sum\limits_{n=1}^N \hbar 
\omega_n S^z_n + \sum\limits_{\vec{k}s} \hbar \omega_k (b^{\dagger}_{\vec{k}s}
b_{\vec{k}s} + {1 / 2}) - i\hbar \sum\limits_{\vec{k}s}
\sum\limits_{n=1}^N [\vec{\mu}_n . \vec{g}_{\vec{k}s} (\vec{r}_n)(S_n^+ + 
S_n^-)b_{\vec{k}s}- h.c.]. \label{hamiltonian} 
\end{eqnarray}
Here $\vec{\mu}_n$ are the transition dipole moments, dependent on the
different atomic positions $\vec{r}_n$ and 
\begin{equation}
S_n^+ = | e_n \rangle \langle g_n|,~ S_n^- = | g_n \rangle \langle e_n|,
\label{dipole}
\end{equation}
are  the dipole raising  and lowering  operators satisfying  the usual
commutation relations and
\begin{equation}
S_n^z = \frac{1}{2}(| e_n \rangle \langle e_n| -  |g_n \rangle \langle g_n|),
\label{energy}
\end{equation}
is    the    energy    operator    of   the    $n$th    atom,    while
$b^{\dagger}_{\vec{k}s}$,   $b_{\vec{k}s}$   are   the  creation   and
annihilation operators  of the field  mode (bath) $\vec{k}s$  with the
wave vector $\vec{k}$, frequency  $\omega_k$ and polarization index $s
= 1,2$ with the system-reservoir (S-R) coupling constant
\begin{equation}
\vec{g}_{\vec{k}s} (\vec{r}_n) = (\frac{\omega_k}{2 \varepsilon \hbar V})^
{1/2} \vec{e}_{\vec{k}s} e^{i \vec{k}.r_n}. \label{coupling}
\end{equation}
Here  $V$  is   the  normalization  volume and $\vec{e}_{\vec{k}s}$ is 
the unit polarization vector of the field.   It  can   be  seen  from
Eq. (\ref{coupling})  that the S-R  coupling constant is  dependent on
the  atomic position  $r_n$. This  leads  to a  number of  interesting
dynamical aspects, as seen below.  From now we will concentrate on the
case of two qubits.  Assuming separable initial conditions, and taking
a trace over  the bath the reduced density matrix  of the qubit system
in the interaction picture and in the usual Born-Markov, rotating wave
approximation (RWA) is obtained as \cite{ft02}
\begin{eqnarray}
\frac{d\rho}{dt} &=& -\frac{i}{\hbar}[H_{\tilde{S}}, \rho] - 
\frac{1}{2} \sum\limits_{i,j=1}^2 \Gamma_{ij}[1+\tilde{N}] 
(\rho S_i^+ S_j^- +  S_i^+ S_j^- \rho - 2 S_j^- \rho  S_i^+)\nonumber\\
&-& \frac{1}{2} \sum\limits_{i,j=1}^2 \Gamma_{ij}\tilde{N} 
(\rho S_i^- S_j^+ +  S_i^- S_j^+ \rho - 2 S_j^+ \rho  S_i^-)
+ \frac{1}{2} \sum\limits_{i,j=1}^2 \Gamma_{ij}\tilde{M} 
(\rho S_i^+ S_j^+ +  S_i^+ S_j^+ \rho - 2 S_j^+ \rho  S_i^+)\nonumber\\
&+& \frac{1}{2} \sum\limits_{i,j=1}^2 \Gamma_{ij}\tilde{M^*} 
(\rho S_i^- S_j^- +  S_i^- S_j^- \rho - 2 S_j^- \rho  S_i^-). \label{red}
\end{eqnarray}
In Eq. (\ref{red}) 
\begin{equation}
\tilde{N} = N_{\rm th}(\cosh^2(r) + \sinh^2(r)) + \sinh^2(r), \label{N} 
\end{equation}
\begin{equation}
\tilde{M} = -\frac{1}{2} \sinh(2r) e^{i\Phi} (2 N_{\rm th} + 1)
\equiv Re^{i\Phi(\omega_0)}, \label{M}
\end{equation}
with 
\begin{equation}
\omega_0 = \frac{\omega_1 + \omega_2}{2},    \label{omega0}
\end{equation}
and
\begin{equation}
N_{\rm th} = {1 \over e^{{\hbar \omega \over k_B T}} - 1}. \label{sqpara}
\end{equation}
Here  $N_{\rm th}$  is the  Planck distribution  giving the  number of
thermal  photons  at  the  frequency  $\omega$  and  $r$,  $\Phi$  are
squeezing parameters.   The analogous case  of a thermal  bath without
squeezing can be obtained from  the above expressions by setting these
squeezing parameters  to zero, while setting the  temperature ($T$) to
zero one recovers the case of the vacuum bath.  Eq. (\ref{red}), for a
single qubit case, can be solved using the Bloch vector formalism (cf.
\cite{bp02},  \cite{bsri06}) and also  in the  superoperator formalism
\cite{bsdiss}. Here the assumption of perfect matching of the squeezed
modes  to the  modes of  the  EMF is  made along  with, the  squeezing
bandwidth  being much larger  than the  atomic linewidths.   Also, the
squeezing carrier frequency is taken to be tuned in resonance with the
atomic frequencies.

In Eq. (\ref{red}),
\begin{equation}
H_{\tilde{S}} = \hbar \sum\limits_{n=1}^2 \omega_n S^z_n + \hbar
\sum^2_{\stackrel{i,j}{(i \neq  j)}} \Omega_{ij} S_i^+ S_j^-, \label{unitary}
\end{equation}
where 
\begin{eqnarray}
\Omega_{ij}   &=&   \frac{3}{4}   \sqrt{\Gamma_i   \Gamma_j}\left[-[1-
(\hat{\mu}.\hat{r}_{ij})^2]\frac{\cos(k_0  r_{ij})}{k_0  r_{ij}} +  [1-
3(\hat{\mu}.\hat{r}_{ij})^2]   \right.   \nonumber\\  &\times&   \left.
[\frac{\sin(k_0    r_{ij})}{(k_0     r_{ij})^2}    +    \frac{\cos(k_0
r_{ij})}{(k_0 r_{ij})^3}]\right].               \label{omegacol}      
\end{eqnarray}
Here $\hat{\mu}  = \hat{\mu}_1  = \hat{\mu}_2$ and  $\hat{r}_{ij}$ are
unit  vectors   along  the   atomic  transition  dipole   moments  and
$\vec{r}_{ij}  = \vec{r}_i  - \vec{r}_j$,  respectively.  Also  $k_0 =
{\omega_0  / c}$,  with $\omega_0$  being as  in  Eq.  (\ref{omega0}),
$r_{ij} = |\vec{r}_{ij}|$.  The wavevector $k_0 = {2\pi / \lambda_0}$,
$\lambda_0$ being  the resonant wavelength, occuring in  the term $k_0
r_{ij}$ sets  up a  length scale into  the problem depending  upon the
ratio  ${r_{ij} /  \lambda_0}$. This  is  thus the  ratio between  the
interatomic  distance  and the  resonant  wavelength,  allowing for  a
discussion  of   the  dynamics  in  two   regimes:  (A).   independent
decoherence  where $k_0.r_{ij}  \sim \frac{r_{ij}}{\lambda_0}  \geq 1$
and    (B).    collective    decoherence   where    $k_0.r_{ij}   \sim
\frac{r_{ij}}{\lambda_0}  \rightarrow 0$. The  case (B)  of collective
decoherence would arise  when the qubits are close  enough for them to
feel the  bath collectively  or when the  bath has a  long correlation
length (set  by the resonant wavelength $\lambda_0$)  in comparison to
the interqubit separation $r_{ij}$.  $\Omega_{ij}$ (\ref{omegacol}) is
a collective coherent effect due to the multi-qubit interaction and is
mediated via the bath through the terms
\begin{equation}
\Gamma_i   =  \frac{\omega^3_i   \mu^2_i}{3   \pi  \varepsilon   \hbar
c^3}. \label{single}
\end{equation}
The  term $\Gamma_i$  is  present  even in  the  case of  single-qubit
dissipative system  bath interaction \cite{bsri06, bsdiss}  and is the
spontaneous emission rate, while
\begin{equation}
\Gamma_{ij} = \Gamma_{ji} = \sqrt{\Gamma_i \Gamma_j} F(k_0 r_{ij}),
\label{gammacol}
\end{equation}
where $i \neq j$ with
\begin{eqnarray}
F(k_0 r_{ij})   &=&   \frac{3}{2} \left[[1-
(\hat{\mu}.\hat{r}_{ij})^2]\frac{\sin(k_0  r_{ij})}{k_0  r_{ij}} +  [1-
3(\hat{\mu}.\hat{r}_{ij})^2]   \right.   \nonumber\\  &\times&   \left.
[\frac{\cos(k_0    r_{ij})}{(k_0     r_{ij})^2}    -    \frac{\sin(k_0
r_{ij})}{(k_0 r_{ij})^3}]\right].               \label{fcol}      
\end{eqnarray}
$\Gamma_{ij}$ (\ref{gammacol}) is the collective incoherent effect due
to the dissipative multi-qubit interaction with the bath. For the case
of identical qubits, as  considered here, $\Omega_{12} = \Omega_{21}$,
$\Gamma_{12} = \Gamma_{21}$ and $\Gamma_1 = \Gamma_2 = \Gamma$.

\section{Dynamics  of  the Two-Qubit  Dissipative  interaction with  a
Vacuum and Squeezed Thermal Bath}

Here  we  present  the   solutions  of  the  density  matrix  equation
(\ref{red})  for the  case of  a two-qubit  system interacting  with a
(A). vacuum bath and (B). squeezed thermal bath. These results will be
of  use   in  the  investigation  of  the   dynamics  of  entanglement
subsequently.

\subsection{Vacuum bath}

Here  we proceed  as in  \cite{ft02}  and obtain  the reduced  density
matrix from Eq.  (\ref{red}), setting  $T$ and bath squeezing to zero,
and  by going  over  to the  dressed  state basis,  of the  collective
two-qubit  dynamics,  obtained  from the  Hamiltonian  $H_{\tilde{S}}$
(\ref{unitary}) as
\begin{eqnarray}
|g \rangle &=& |g_1 \rangle |g_2 \rangle, \nonumber\\
|s \rangle &=& \frac{1}{\sqrt{2}} (|e_1 \rangle |g_2 \rangle + 
|g_1 \rangle |e_2 \rangle), \nonumber\\
|a \rangle &=& \frac{1}{\sqrt{2}} (|e_1 \rangle |g_2 \rangle - 
|g_1 \rangle |e_2 \rangle), \nonumber\\
|e \rangle &=& |e_1 \rangle |e_2 \rangle, \label{dressed}
\end{eqnarray}
with the corresponding eigenvalues being $E_g = -\hbar \omega_0$, $E_s
=  \hbar \Omega_{12}$,  $E_a =  -\hbar \Omega_{12}$  and $E_g  = \hbar
\omega_0$.   The reduced  density matrix  in the  dressed  state basis
(\ref{dressed}) can be obtained from Eq. (\ref{red}) as
\begin{equation}
\frac{d\rho}{dt} = -\frac{i}{\hbar}[H_{as}, \rho] + (\frac{d\rho}{dt})_s 
+ (\frac{d\rho}{dt})_a,  \label{redvac}
\end{equation}
where
\begin{equation}
H_{as} = \hbar \left[\omega_0 (|e \rangle \langle e| - 
|g \rangle \langle g|) + \Omega_{12} (|s \rangle \langle s| - 
|a \rangle \langle a|) \right], \label{has} 
\end{equation}
\begin{eqnarray}
(\frac{d\rho}{dt})_s  &=& - \frac{1}{2} (\Gamma + \Gamma_{12}) \left[
(|e \rangle \langle e| + |s \rangle \langle s|) \rho +
\rho (|e \rangle \langle e| + |s \rangle \langle s|) \right. \nonumber\\
&-& \left. 2  (|s \rangle \langle e| + |g \rangle \langle s|) \rho
(|e \rangle \langle s| + |s \rangle \langle g|) \right], \label{rhos}
\end{eqnarray}
and
\begin{eqnarray}
(\frac{d\rho}{dt})_a  &=& - \frac{1}{2} (\Gamma - \Gamma_{12}) \left[
(|e \rangle \langle e| + |a \rangle \langle a|) \rho +
\rho (|e \rangle \langle e| + |a \rangle \langle a|) \right. \nonumber\\
&-& \left. 2  (|a \rangle \langle e| - |g \rangle \langle a|) \rho
(|e \rangle \langle a| - |a \rangle \langle g|) \right]. \label{rhoa}
\end{eqnarray}
From the dressed state basis  (\ref{dressed}), it can be seen that the
two-qubit  problem  can be  thought  of  as  an equivalent  four-level
system. For the case where ${r_{ij} / \lambda_0} \rightarrow 0$, i.e.,
when  the interatomic  separation is  much smaller  than  the resonant
wavelength,  constituting  the Dicke  model  \cite{dicke}, $({d\rho  /
dt})_a  = 0$  and  the  problem reduces  to  an effective  three-level
system.  The  Eq. (\ref{redvac})  can be solved  to yield  the various
density matrix elements as follows:
\begin{equation}
\rho_{ee}  (t)   =  e^{-2\Gamma t} \rho_{ee}  (0), \label{rhoeevac}
\end{equation}
\begin{equation}
\rho_{ss}  (t)   =  e^{-(\Gamma  +  \Gamma_{12})t}   \rho_{ss}  (0)  +
\frac{(\Gamma + \Gamma_{12})}{(\Gamma - \Gamma_{12})} (1 - e^{-(\Gamma
- \Gamma_{12})t}   )e^{-(\Gamma   +   \Gamma_{12})t}  \rho_{ee}   (0),
\label{rhossvac}
\end{equation}
\begin{equation}
\rho_{aa}  (t)   =  e^{-(\Gamma  -  \Gamma_{12})t}   \rho_{aa}  (0)  +
\frac{(\Gamma - \Gamma_{12})}{(\Gamma + \Gamma_{12})} (1 - e^{-(\Gamma
+ \Gamma_{12})t}   )e^{-(\Gamma   -   \Gamma_{12})t}  \rho_{ee}   (0),
\label{rhoaavac}
\end{equation}
\begin{eqnarray}
\rho_{gg}  (t) &=& \rho_{gg}  (0) + (1 - e^{-(\Gamma
+ \Gamma_{12})t}) \rho_{ss}  (0) + (1 - e^{-(\Gamma
- \Gamma_{12})t}) \rho_{aa}  (0) \nonumber\\ &+& \left[
\frac{(\Gamma + \Gamma_{12})}{2 \Gamma} \Big\{1 - \frac{2}{(\Gamma -
\Gamma_{12})} [\frac{(\Gamma + \Gamma_{12})}{2}(1- e^{-(\Gamma
- \Gamma_{12})t}) + \frac{(\Gamma - \Gamma_{12})}{2}]  
e^{-(\Gamma  +  \Gamma_{12})t}\Big\} \right. \nonumber\\ &+& \left.
\frac{(\Gamma - \Gamma_{12})}{(\Gamma + \Gamma_{12})} \Big\{
(1 - e^{-(\Gamma - \Gamma_{12})t}) -  \frac{(\Gamma - \Gamma_{12})}{2 \Gamma}
(1 - e^{-2 \Gamma t}) \Big\} \right] \rho_{ee}   (0). \label{rhoggvac}
\end{eqnarray}
The Eqs. (\ref{rhoeevac}) to (\ref{rhoggvac}) give the dynamics of the
population of the two-qubit system  interacting with a vacuum bath and
$\rho_{ee}  (t) +  \rho_{ss} (t)  + \rho_{aa}  (t) +  \rho_{gg}  (t) =
\rho_{ee} (0)  + \rho_{ss} (0) +  \rho_{aa} (0) +  \rho_{gg} (0)$. The
off- diagonal terms of the density matrix are:
\begin{eqnarray}
\rho_{es} (t) &=& e^{-i(\omega_0 - \Omega_{12})t} e^{-\frac{1}{2}(3 \Gamma
+ \Gamma_{12})t} \rho_{es} (0), \nonumber\\
\rho_{se} (t) &=& \rho^*_{es} (t), \label{rhoesvac}
\end{eqnarray}
\begin{eqnarray}
\rho_{eg} (t) &=& e^{-i2\omega_0 t} e^{- \Gamma
t} \rho_{eg} (0), \nonumber\\
\rho_{ge} (t) &=& \rho^*_{eg} (t), \label{rhoegvac}
\end{eqnarray}
\begin{eqnarray}
\rho_{ea} (t) &=& e^{-i(\omega_0 + \Omega_{12})t} e^{-\frac{1}{2}(3 \Gamma
- \Gamma_{12})t} \rho_{ea} (0), \nonumber\\
\rho_{ae} (t) &=& \rho^*_{ea} (t), \label{rhoeavac}
\end{eqnarray}
\begin{eqnarray}
\rho_{sa} (t) &=& e^{-i2\Omega_{12} t} e^{- \Gamma
t} \rho_{sa} (0), \nonumber\\
\rho_{as} (t) &=& \rho^*_{sa} (t), \label{rhosavac}
\end{eqnarray}
\begin{eqnarray}
\rho_{ag}     (t)      &=&     e^{-i(\omega_0     -     \Omega_{12})t}
e^{-\frac{1}{2}(\Gamma - \Gamma_{12})t}  \rho_{ag} (0) - \frac{(\Gamma
  - \Gamma_{12})}  {(\Gamma^2  +  4 \Omega^2_{12})}  e^{-i(\omega_0  -
  \Omega_{12})t} \nonumber\\ &\times&
e^{-\frac{1}{2}(\Gamma - \Gamma_{12})t} \left[ 
2 \Omega_{12} e^{- \Gamma t} \sin(2 \Omega_{12}t)
+ \Gamma (1 - e^{- \Gamma t} \cos(2 \Omega_{12}t))\right]  \rho_{ea} (0)
\nonumber\\ &+& i  \frac{(\Gamma
  - \Gamma_{12})}  {(\Gamma^2  +  4 \Omega^2_{12})} 
 e^{-i(\omega_0  -
  \Omega_{12})t} e^{-\frac{1}{2}(\Gamma - \Gamma_{12})t} \left[
 2 \Omega_{12}(1 - e^{- \Gamma t} \cos(2 \Omega_{12}t)) \right.\nonumber\\
&-& \left. \Gamma
 e^{- \Gamma t} \sin(2 \Omega_{12}t) \right] \rho_{ea} (0), \nonumber\\
\rho_{ga} (t) &=& \rho^*_{ag} (t), \label{rhoagvac}  
\end{eqnarray}
\begin{eqnarray}
\rho_{sg}     (t)      &=&     e^{-i(\omega_0     +     \Omega_{12})t}
e^{-\frac{1}{2}(\Gamma + \Gamma_{12})t}  \rho_{sg} (0) + \frac{(\Gamma
  + \Gamma_{12})}  {(\Gamma^2  +  4 \Omega^2_{12})}  e^{-i(\omega_0  +
  \Omega_{12})t} \nonumber\\ &\times&
e^{-\frac{1}{2}(\Gamma + \Gamma_{12})t} \left[ 
2 \Omega_{12} e^{- \Gamma t} \sin(2 \Omega_{12}t)
+ \Gamma (1 - e^{- \Gamma t} \cos(2 \Omega_{12}t))\right]  \rho_{es} (0)
\nonumber\\ &+& i  \frac{(\Gamma
  + \Gamma_{12})}  {(\Gamma^2  +  4 \Omega^2_{12})} 
 e^{-i(\omega_0  +
  \Omega_{12})t} e^{-\frac{1}{2}(\Gamma + \Gamma_{12})t} \left[
 2 \Omega_{12}(1 - e^{- \Gamma t} \cos(2 \Omega_{12}t)) \right.\nonumber\\
&-& \left. \Gamma
 e^{- \Gamma t} \sin(2 \Omega_{12}t) \right] \rho_{es} (0), \nonumber\\
\rho_{gs} (t) &=& \rho^*_{sg} (t). \label{rhosgvac}  
\end{eqnarray}
The sixteen density matrix elements given by Eqs.  (\ref{rhoeevac}) to
(\ref{rhosgvac}) completely  solve the master  equation (\ref{redvac})
describing the  dynamics of the two-qubit system  interacting with the
vacuum  bath.  For  the  case of  the  Dicke model,  where ${r_{ij}  /
  \lambda_0} \rightarrow 0$, i.e.,  when the interatomic separation is
much smaller than the resonant wavelength, $ ({d\rho / dt})_a = 0$ and
the solution of the effective  three-level system can be extracted out
of the above equations. The  conditions under which the Dicke model is
obtained is  analogous to the  case of collective decoherence  for the
case of two-qubit interaction with  a bath via a quantum nondemolition
interaction (QND) \cite{brs1}. There it was found that for the case of
interaction  with a  thermal (and  also a  vacuum) bath,  the subspace
spanned    by     $\{|e_1,g_2\rangle,    |g_1,e_2\rangle\}$    is    a
decoherence-free   subspace,  implying   that   the  matrix   elements
$\rho_{e_1,g_2;   e_1,g_2},   \rho_{g_1,e_2;g_1,e_2}$  $\rho_{e_1,g_2;
  g_1,e_2}$ and  $\rho_{g_1,e_2;e_1,g_2}$ remain invariant  inspite of
the    system's   interaction    with   a    bath.     However,   from
Eqs. (\ref{rhoeevac})--(\ref{rhosgvac})  it is clear that  none of the
matrix elements  is invariant  as a function  of time,  reflecting the
greater complexity of the dissipative interaction.
 
\subsection{Squeezed thermal bath}

Here we consider the  two-qubit dynamics resulting from an interaction
with a squeezed thermal bath, i.e., make use of Eq.  (\ref{red}).  The
equations  of the  reduced  density matrix  (\ref{red})  taken in  the
two-qubit  dressed state  basis (\ref{dressed})  are not  all mutually
coupled, but divide into four irreducible blocks A, B, C, D,
thereby reducing the task from an evaluation of fifteen coupled linear
differential  equations   to  that  of  a  maximum   of  four  coupled
equations. Thus we have:

\underline{Block A}:
\begin{eqnarray} 
\dot{\rho}_{ee} (t) &=& -2 \Gamma (\tilde{N} + 1) \rho_{ee} (t) +
\tilde{N} \Big\{(\Gamma + \Gamma_{12}) \rho_{ss} (t) + 
(\Gamma - \Gamma_{12}) \rho_{aa} (t) \Big\} + \Gamma_{12} |\tilde{M}|
\rho_u (t), \nonumber\\
\dot{\rho}_{ss} (t) &=& -(\Gamma + \Gamma_{12}) \Big\{-\tilde{N} +
(1 + 3 \tilde{N}) \rho_{ss} (t) - \rho_{ee} (t) + \tilde{N} \rho_{aa} (t) +
|\tilde{M}| \rho_u (t) \Big\}, \nonumber\\
\dot{\rho}_{aa} (t) &=& (\Gamma - \Gamma_{12}) \Big\{\tilde{N} -
(1 + 3 \tilde{N}) \rho_{aa} (t) + \rho_{ee} (t) - \tilde{N} \rho_{ss} (t) +
|\tilde{M}| \rho_u (t) \Big\}, \nonumber\\ 
\dot{\rho}_u (t) &=& 2 |\tilde{M}| \Gamma_{12} - 4 \omega_0 |\rho_{ge}|
\sin(\Phi + \chi)- (2 \tilde{N} + 1) \Gamma
\rho_u - 2 |\tilde{M}| \Big\{(\Gamma + 2 \Gamma_{12}) \rho_{ss} (t) - 
(\Gamma - 2 \Gamma_{12}) \rho_{aa} (t) \Big\}. \label{population}
\end{eqnarray}
Here 
\begin{equation}
\rho_u (t) = e^{i \Phi} \rho_{ge} (t) + h.c., \label{rhou}
\end{equation}
$\rho_{ge}   =  |\rho_{ge}|   e^{i  \chi}$   and  $\Phi$   is   as  in
Eq. (\ref{M}). Also $\rho_{gg} (t)  = 1- \rho_{aa} (t) - \rho_{ss} (t)
- \rho_{ee}  (t)$.  The Eqs.  (\ref{population}) give  the dynamics of
the  population of the  two-qubit system  interacting with  a squeezed
thermal bath  while the off-diagonal terms,  given by the  Blocks B, C
and D are:

\underline{Block B}:
\begin{eqnarray}
\dot{\rho}_{es} (t) &=& -i (\omega_0 - \Omega_{12})) \rho_{es} (t) -\frac{1}{2}
\Big\{(3 \Gamma + \Gamma_{12}) + 2 \tilde{N} (2 \Gamma + \Gamma_{12})\Big\}
\rho_{es} (t) + \tilde{N} (\Gamma + \Gamma_{12}) \rho_{sg} (t) 
\nonumber\\ &+&
\tilde{M} \Gamma_{12} \rho_{gs} (t) -
\tilde{M} (\Gamma + \Gamma_{12}) \rho_{se} (t), \nonumber\\
\dot{\rho}_{se} (t) &=& \dot{\rho^*}_{es} (t), \nonumber\\
\dot{\rho}_{gs} (t) &=& -i (\omega_0 - \Omega_{12})) \rho_{gs} (t) -\frac{1}{2}
\Big\{(\Gamma + \Gamma_{12}) + 2 \tilde{N} (2 \Gamma + \Gamma_{12})\Big\}
\rho_{gs} (t) + (1 + \tilde{N}) (\Gamma + \Gamma_{12}) \rho_{se} (t) 
\nonumber\\ &+&
\tilde{M^*} \Gamma_{12} \rho_{es} (t) -
\tilde{M^*} (\Gamma + \Gamma_{12}) \rho_{sg} (t), \nonumber\\
\dot{\rho}_{sg} (t) &=& \dot{\rho^*}_{gs} (t). \label{blockB}
\end{eqnarray}

\underline{Block C}:
\begin{eqnarray}
\dot{\rho}_{as} (t) &=& i  2  \Omega_{12} \rho_{as} (t) - \Gamma (1 + 
2 \tilde{N})  \rho_{as} (t), \nonumber\\
\dot{\rho}_{sa} (t) &=& \dot{\rho^*}_{as} (t). \label{blockC}
\end{eqnarray}

\underline{Block D}:
\begin{eqnarray}
\dot{\rho}_{ea} (t) &=& -i (\omega_0 + \Omega_{12})) \rho_{ea} (t) -\frac{1}{2}
\Big\{(3 \Gamma - \Gamma_{12}) + 2 \tilde{N} (2 \Gamma - \Gamma_{12})\Big\}
\rho_{ea} (t) - \tilde{N} (\Gamma - \Gamma_{12}) \rho_{ag} (t) 
\nonumber\\ &+&
\tilde{M} \Gamma_{12} \rho_{ga} (t) -
\tilde{M} (\Gamma - \Gamma_{12}) \rho_{ae} (t), \nonumber\\
\dot{\rho}_{ae} (t) &=& \dot{\rho^*}_{ea} (t), \nonumber\\
\dot{\rho}_{ga} (t) &=& i (\omega_0 - \Omega_{12})) \rho_{ga} (t) -\frac{1}{2}
\Big\{(\Gamma - \Gamma_{12}) + 2 \tilde{N} (2 \Gamma - \Gamma_{12})\Big\}
\rho_{ga} (t) - (1 + \tilde{N}) (\Gamma - \Gamma_{12}) \rho_{ae} (t) 
\nonumber\\ &+&
\tilde{M^*} \Gamma_{12} \rho_{ea} (t) -
\tilde{M^*} (\Gamma - \Gamma_{12}) \rho_{ag} (t), \nonumber\\
\dot{\rho}_{ag} (t) &=& \dot{\rho^*}_{ga} (t). \label{blockD}
\end{eqnarray}

The Eqs. (\ref{blockC}) can be trivially solved to yield
\begin{eqnarray}
\rho_{as} (t) &=& e^{[i 2 \Omega_{12} - \Gamma(1 + 2 \tilde{N})]t} 
\rho_{as} (0) \nonumber\\ \rho_{sa} (t) &=& \rho^*_{as} (t), \label{blockC1} 
\end{eqnarray}
while  the  blocks  A,  B   and  D  consist  of  four  linear  coupled
differential equations which can be written in matrix form as:
\begin{equation}
\dot{P} (t) = -Q P(t) + W, \label{matrix}
\end{equation}
where $Q$  is a time-independent $4  \times 4$ matrix  and $P(t)$, $W$
are  $4 \times  1$  column  vectors. The  Eq.  (\ref{matrix}) has  the
general solution
\begin{equation}
P(t) = (V e^{-D t} V^{-1}) P(0) + V D^{-1} (1 - e^{-D t}) V^{-1} W. 
\label{matrix2}
\end{equation}
Here $V$ is the vector composed of the eigenvectors of the matrix $Q$,
while   $D$  is   composed   of  its   eigenvalues.    We  solve   Eq.
(\ref{matrix2})   by  numerically   obtaining   the  eigenvalues   and
eigenvectors of the matrix $Q$ for the Blocks A, B, C, and D.
\begin{figure}
\subfigure[]{\includegraphics[width=7.0cm]{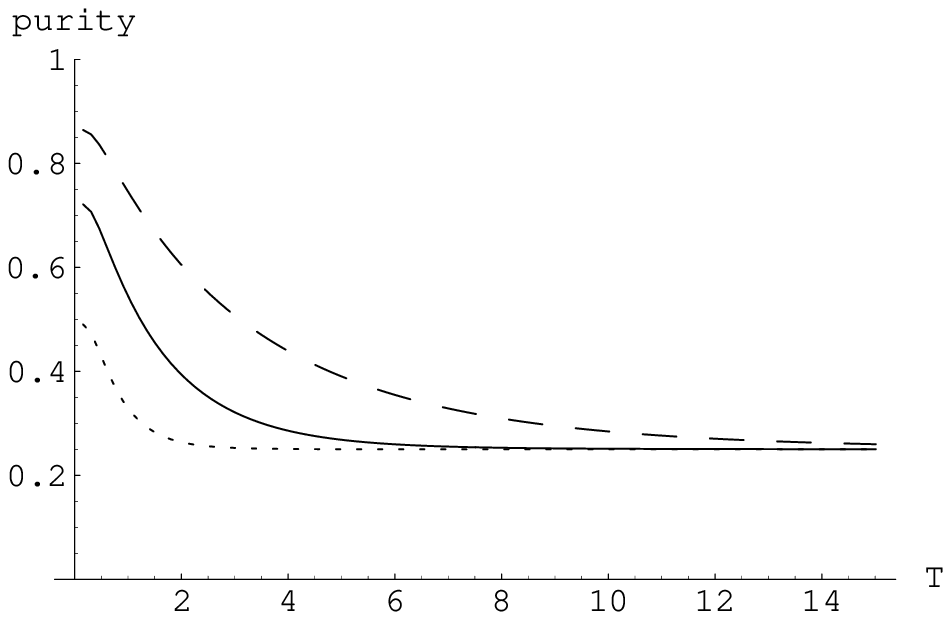}} 
\hfill
\subfigure[]{\includegraphics[width=7.0cm]{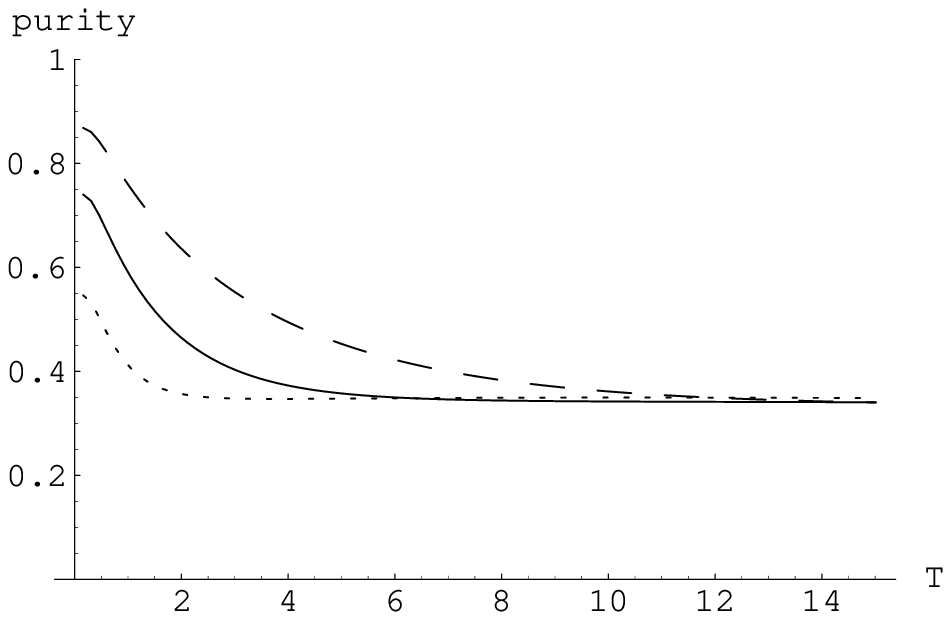}}
\caption{  Purity as  a function  of temperature  $T$ (in  units where
  $\hbar\equiv  k_B=1$)  for (a)  the  independent decoherence  model,
  where ${r_{ij}  / \lambda_0}$ (\ref{omegacol})  is $\geq 1$  and (b)
  the  collective  decoherence  model,  where ${r_{ij}  /  \lambda_0}$
  (\ref{omegacol})  is  $\approx  0$.    Here  with  $r_{12}$  is  the
  inter-qubit  distance.   The large-dashed,  bold  and dotted  curves
  correspond to  evolution time  $t=1.0$ and bath  squeezing parameter
  (\ref{N},  \ref{M}) $r=-0.5$, $1.0$  and $1.5$,  respectively.  Here
  and in  all the subsequent  figures, the squeezing  parameter $\Phi$
  (\ref{M}) is set equal  to zero.  Also $\omega_0$ (\ref{omega0}) and
  the bath parameter $\Gamma$ (\ref{single}), are set equal to 1.0 and
  0.05, respectively. All the inter-qubit distances are defined on the
  scale  of the  resonant wavelength  coming from  the  wavevector $k$
  (\ref{coupling}) as a result  of the position dependent couplings of
  the qubits with the bath.   In figure (a) related to the independent
  decoherence model, $kr_{12}$ is set equal to 1.5 while in figure (b)
  related to the collective  decoherence model, $kr_{12}$ is set equal
  to 0.08.}
\label{fig:purity}
\end{figure}

Figures  (\ref{fig:purity} (a)),  (b) depict  the behavior  of purity,
defined here  as ${\rm Tr}(\rho^2  (t))$ for $\rho(t)$ as  obtained in
this  subsection  for  the   independent  ($k_0.r_{ij}  \neq  0$)  and
collective    ($k_0.r_{ij}   \rightarrow   0$)    decoherence   model,
respectively, as a  function of temperature $T$ for  an evolution time
$t$ and bath  squeezing $r$ (\ref{N}, \ref{M}). In  all the figures in
this  article, we  consider  the initial  state  of one  qubit in  the
excited state  $|e_1 \rangle$  and the other  in the ground  state $|g_2
\rangle$, i.e.,  $|e_1\rangle|g_2\rangle$ and $\hat{\mu}.\hat{r}_{ij}$
(\ref{fcol}) is equal to zero.  It  can be seen that with the increase
in temperature, as also evolution time $t$ and bath squeezing $r$, the
system becomes more mixed and hence loses its purity.

\section{Entanglement Analysis}

In this section, we will  study the development of entanglement in the
two qubit system,  both for the independent as  well as the collective
decoherence  model interacting with  a squeezed  thermal bath.  A well
known measure of MSE is the concurrence \cite{ww98} defined as
\begin{equation} 
{\cal C} = \max(0, \sqrt{\lambda_1} - \sqrt{\lambda_2} - \sqrt{\lambda_3}-
\sqrt{\lambda_4}), \label{concur}
\end{equation}
where ${\lambda_i}$ are the eigenvalues of the matrix 
\begin{equation}
R = \rho \tilde{\rho}, \label{concur1}
\end{equation}
with $\tilde{\rho} = \sigma_y \otimes \sigma_y \rho^* \sigma_y \otimes
\sigma_y$ and $\sigma_y$ is the usual Pauli matrix. ${\cal C}$ is zero
for unentangled  states and one for maximally  entangled states. Since
the  reduced dynamics  of the  two-qubit  system was  obtained in  the
dressed state basis  (\ref{dressed}), which contains entangled states,
in order  to remove spurious  entanglement coming from the  basis, for
the entanglement  analysis we  rotate the density  operator back  to a
separable basis  by means of  a Hadamard transformation acting  in the
subspace spanned by $\{|a\rangle,|s\rangle\}$, i.e., the tensor sum of
a Hadamard in this subspace  and an identity operation in the subspace
spanned   by   $\{|e\rangle,|g\rangle\}$,   i.e.,   $H_{(as)}   \oplus
I_{(eg)}$.

\begin{figure}
\subfigure[]{\includegraphics[width=7.0cm]{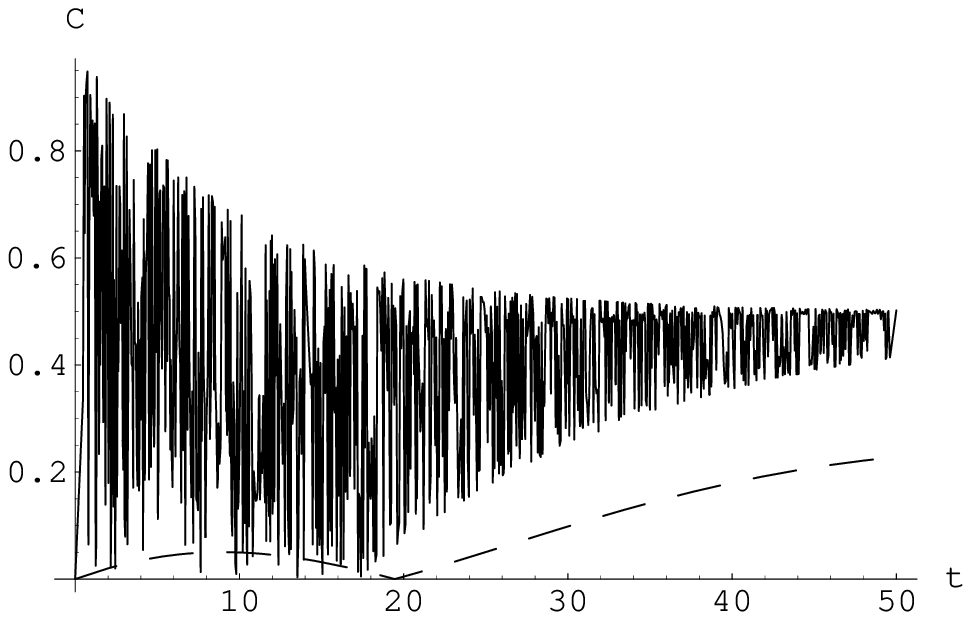}} 
\hfill
\subfigure[]{\includegraphics[width=7.0cm]{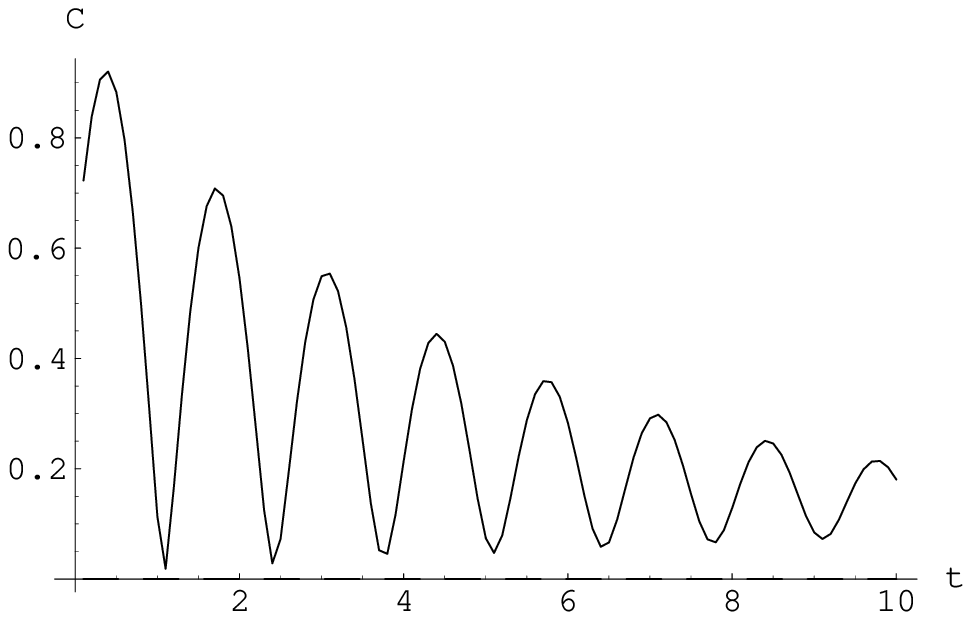}}
\caption{Concurrence ${\cal  C}$ (\ref{concur}) as a  function of time
  of evolution $t$.  Figure (a) deals with the case of vacuum bath ($T
  = r = 0$), while  figure (b) considers concurrence in the two-qubit
  system interacting  with a squeezed thermal bath,  for a temperature
  $T  = 1$  and and  bath squeezing  parameter $r$  (\ref{N}, \ref{M})
  equal  to $0.1$.  In  both the  figures the  bold curve  depicts the
  collective  decoherence  model  ($kr_{12}=0.05$), while  the  dashed
  curve represents the  independent decoherence model ($kr_{12}=1.1$).
  In  figure (b)  for  the  given settings,  the  concurrence for  the
  independent decoherence model is negligible and is thus not seen.}
\label{fig:concuTim}
\end{figure}

\begin{figure}
\subfigure[]{\includegraphics[width=7.0cm]{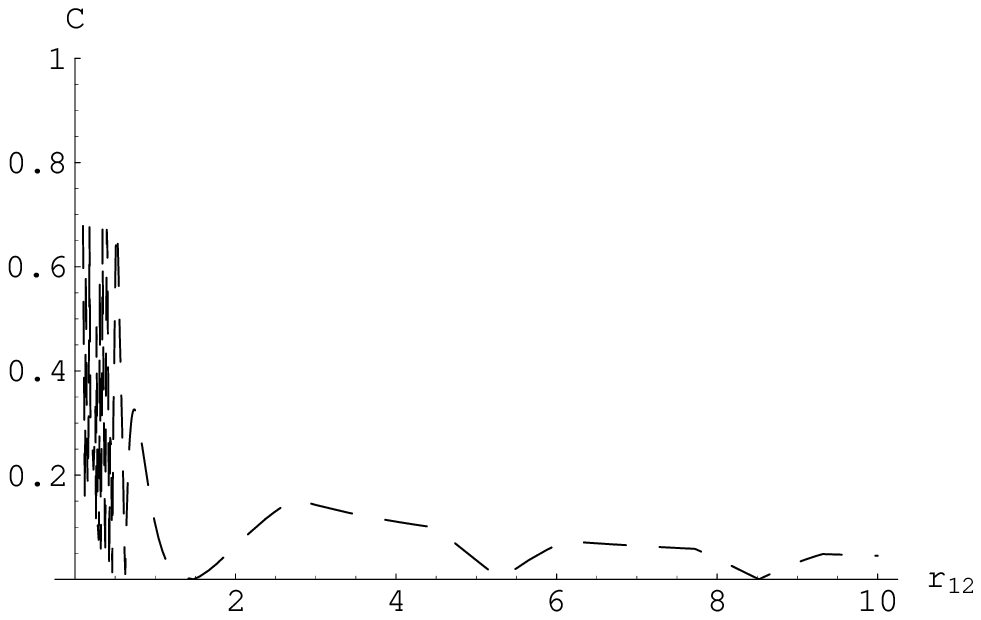}} 
\hfill
\subfigure[]{\includegraphics[width=7.0cm]{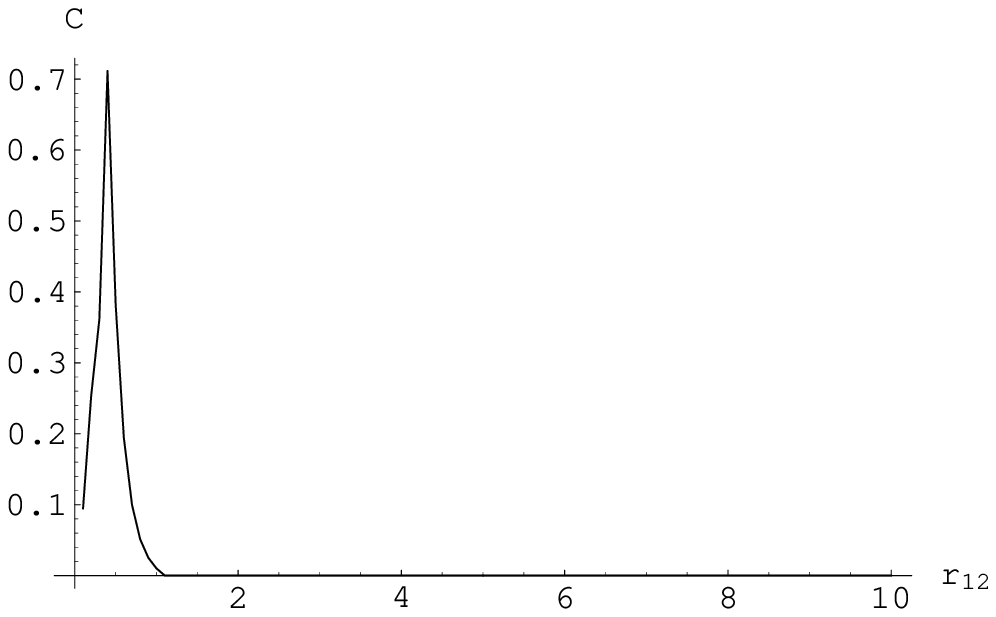}}
\caption{Concurrence  ${\cal   C}$  (\ref{concur})  with   respect  to
  inter-qubit distance  $r_{12}$.  Figure (a)  deals with the  case of
  vacuum bath ($T = r = 0$), while figure (b) considers concurrence in
  the two-qubit  system interacting with a squeezed  thermal bath, for
  $T =  1$, evolution time  $t = 1$  and bath squeezing  parameter $r$
  (\ref{N}, \ref{M})  equal to $0.1$.   In figure (a)  the oscillatory
  behavior of  concurrence is  stronger in the  collective decoherence
  regime,  in  comparison  with  the  independent  decoherence  regime
  ($kr_{12}\geq  1$).   In  figure  (b),  the effect  of  finite  bath
  squeezing and $T$ has the effect of diminishing the concurrence to a
  great  extent  in comparison  to  the  vacuum  bath case.  Here  the
  concurrence for the independent decoherence regime is negligible, in
  agreement with the previous figure.}
\label{fig:concuLength}
\end{figure}

Here we study concurrence for  the two-qubit system interacting with a
squeezed vacuum  bath.  In figure  (\ref{fig:concuTim}) concurrence is
plotted for  the initial  state $|e_1\rangle|g_2\rangle$ for  both the
independent    as    well     as    collective    dynamics.     Figure
(\ref{fig:concuLength})  depicts the behavior  of concurrence  for the
same  initial   state  with   respect  to  the   inter-qubit  distance
$r_{12}$.  It is  clearly seen  that  the buildup  of entanglement  is
greater for  the collective dynamics when compared  to the independent
one.

\begin{figure}
\subfigure[]{\includegraphics[width=7.0cm]{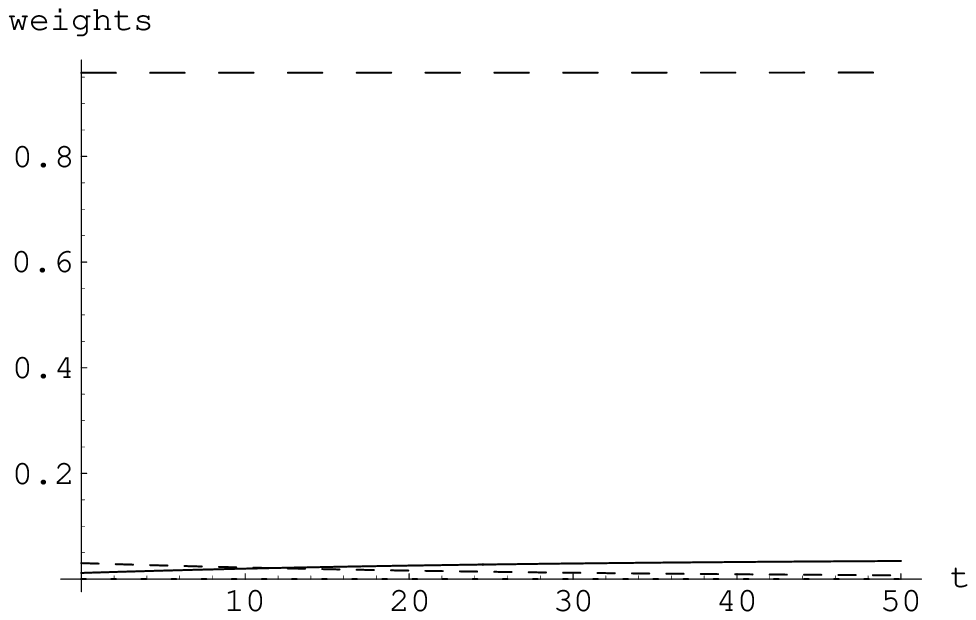}} 
\hfill
\subfigure[]{\includegraphics[width=7.0cm]{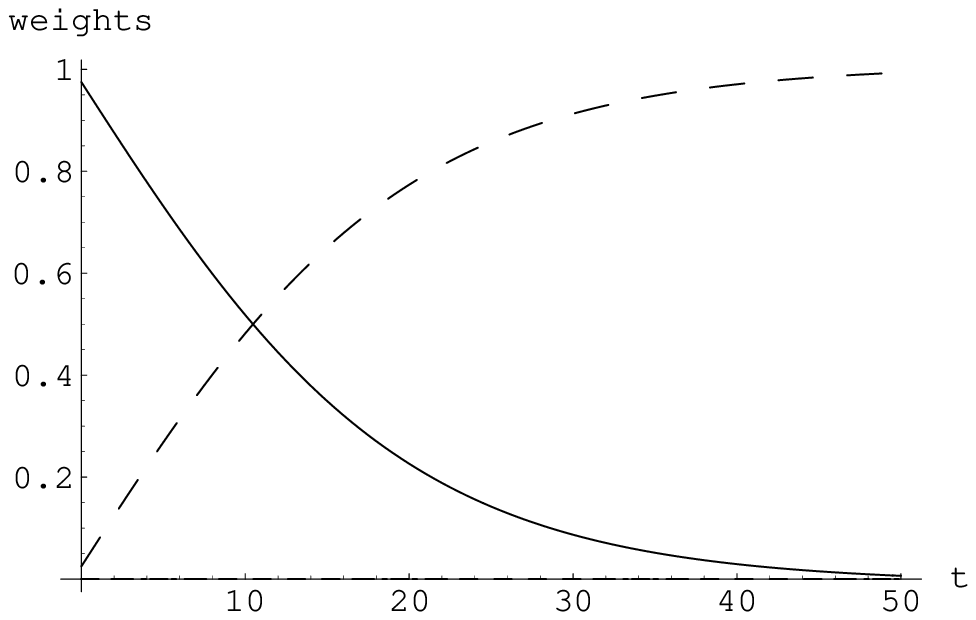}}
\caption{The weights (\ref{PDF}) as  a function of evolution time $t$,
  for  the case  of an  interaction  with an  unsqueezed vacuum  bath.
  Figure  (a)  refers  to  the  independent  decoherence  model,  with
  $kr_{12}  = 1.5$  and  (b) the  collective  decoherence model,  with
  $kr_{12} = 0.08$. In both the figures, the bold curve corresponds to
  the  weight  $\omega_1$, while  the  large-dashed, small-dashed  and
  dotted curves  correspond to the weights  $\omega_2$, $\omega_3$ and
  $\omega_4$,   respectively.   In  both   the  figures,   the  weight
  $\omega_4$ is negligible and hence is not seen.}
\label{fig:weightsvac}
\end{figure}

Now we take  up the issue of entanglement from  the perspective of the
PDF as in Eq.  (\ref{PDF}).  In figures (\ref{fig:weightsvac} (a)) and
(b),  we  plot  the  weights $\omega_1$,  $\omega_2$,  $\omega_3$  and
$\omega_4$ (\ref{PDF}) of the entanglement densities of the projection
operators of  the various subspaces  which span the two  qubit Hilbert
space with respect  to the evolution time $t$  for the independent and
collective  decoherence  models,  respectively,  for the  case  of  an
interaction with  an unsqueezed vacuum bath.  Since  $\omega_1$ is the
weight of  the one dimensional projection, representing  a pure state,
and $\omega_4$ that  of the maximally mixed state  with $\omega_2$ and
$\omega_3$ being intermediary, these plots depict the variation in the
contribution  of the  various  subspaces to  the  entanglement of  the
two-qubit system  as $t$  increases. From the  figures it can  be seen
that in the case of  the independent decoherence model, as depicted in
figure (\ref{fig:weightsvac} (a)), the weight $\omega_2$ dominates the
other weights and remains almost constant, while the remaining weights
are much lower and their increase  is very small compared to it.  This
is in  contrast to the  collective decoherence model, wherein  we find
the weight $\omega_1$ decreases  while $\omega_2$ increases with time.
Since the weight $\omega_1$  is indicative of pure state entanglement,
it is  evident from the figures  that the entanglement  content in the
collective decoherence  model is higher  than that in  the independent
decoherence model.  Also since the weight $\omega_2$, representing the
weight of the two-dimensional projection operator corresponding to the
PDF ${\cal  P}_2 ({\cal  E})$, has a  rich entanglement  structure, we
come to the conclusion that  for the case of the two-qubit interaction
with a vacuum bath, the entanglement  is preserved in the system for a
long time.  In both the  figures, the weight $\omega_4$, indicative of
a  completely   mixed  state,  is  negligible  over   the  time  range
considered.
 
\begin{figure}
\subfigure[]{\includegraphics[width=7.0cm]{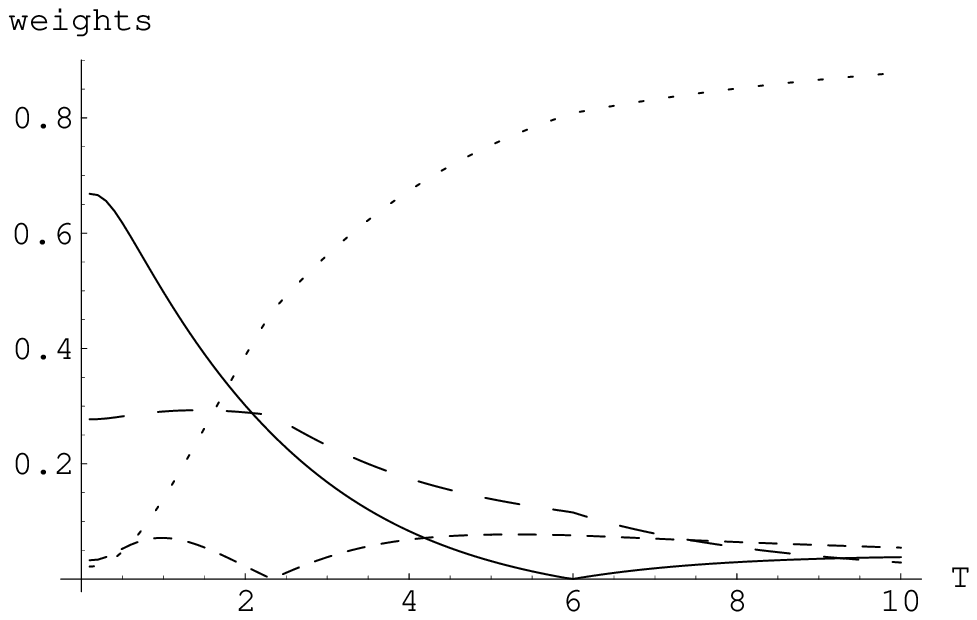}} 
\hfill
\subfigure[]{\includegraphics[width=7.0cm]{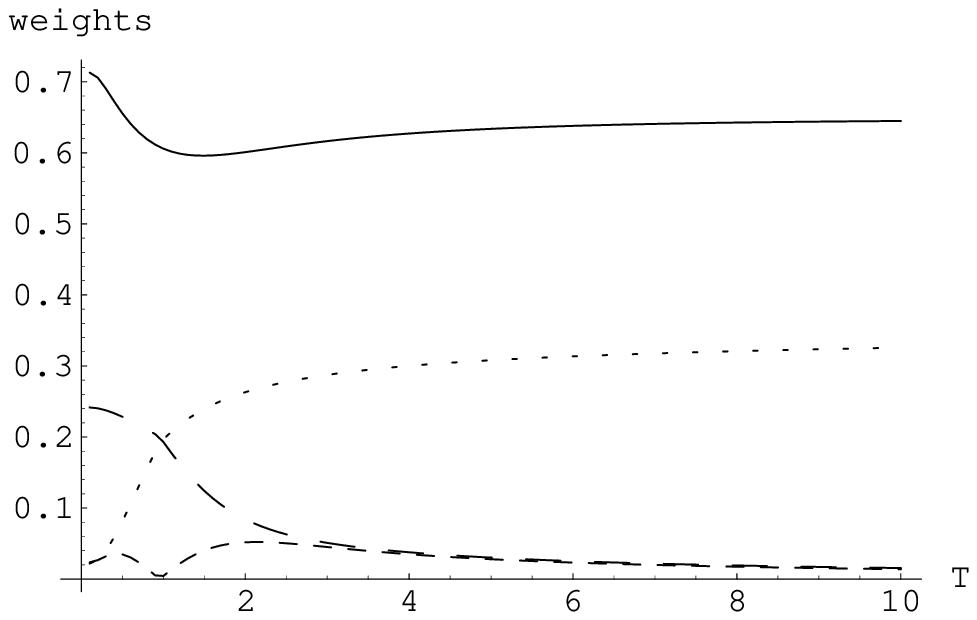}}
\caption{The weights (\ref{PDF}) as a function of temperature $T$, for
  the  case of  an interaction  with a  squeezed thermal  bath  for an
  evolution time  $t = 5$  and bath squeezing parameter  $r$ (\ref{N},
  \ref{M})  equal to  $0.5$.   Figure (a)  refers  to the  independent
  decoherence  model, with  $kr_{12}  = 1.5$  and  (b) the  collective
  decoherence model, with  $kr_{12} = 0.08$. In both  the figures, the
  bold  curve   corresponds  to  the  weight   $\omega_1$,  while  the
  large-dashed,  small-dashed  and  dotted  curves correspond  to  the
  weights $\omega_2$, $\omega_3$ and $\omega_4$, respectively.}
\label{fig:weights}
\end{figure}

In  figures  (\ref{fig:weights} (a))  and  (b),  we  plot the  weights
$\omega_1$, $\omega_2$,  $\omega_3$ and $\omega_4$  (\ref{PDF}) of the
entanglement  densities of  the  projection operators  of the  various
subspaces which span  the two qubit Hilbert space  with respect to $T$
for the  independent and collective  decoherence models, respectively.
These  plots thus  depict the  variation  in the  contribution of  the
various subspaces to  the entanglement of the two  qubit system as $T$
increases.  From  figure (\ref{fig:weights} (b)),  we can see  that in
the  collective decoherence  regime, the  weight  $\omega_1$ initially
falls  and then  stabilizes  around a  finite  value while  $\omega_4$
rises, but remains well below the value of $\omega_1$, indicating that
for the  collective decoherence model,  under the given  settings, the
system maintains a finite value  of entanglement.  This feature is not
observed  in the figure  (\ref{fig:weights} (a)),  where for  the same
settings,  $\omega_1$  decreases  to  zero,  while  $\omega_4$  rises,
thereby indicating a loss of purity and destruction of entanglement.

\begin{figure}
\includegraphics{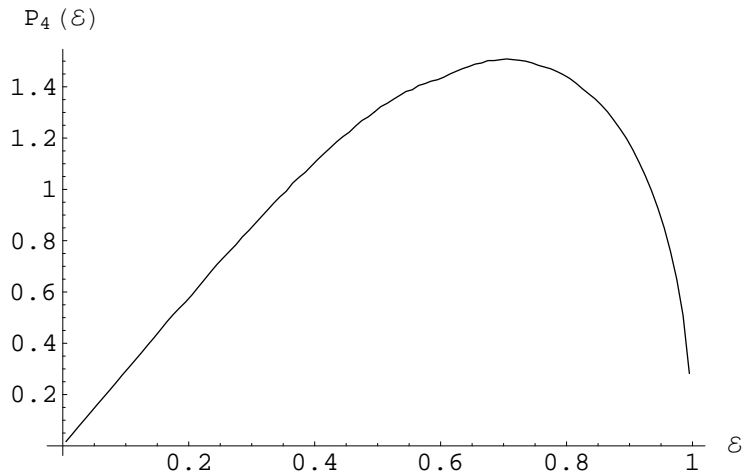}
\caption{The PDF for the four dimensional projection spanning the full
Hilbert space ${\cal H}(\Pi_4)$.}
\label{fig:p4}
\end{figure}

As  explained  in  Section  II,  the characterization  of  MSE  for  a
two-qubit  system involves  the  density function  of four  projection
operators, $\Pi_1$,  $\Pi_2$, $\Pi_3$, $\Pi_4$,  corresponding to one,
two, three, and four dimensional projections, respectively. These will
be represented  here as  ${\cal P}_1 ({\cal  E})$, ${\cal  P}_2 ({\cal
  E})$,  ${\cal  P}_3  ({\cal   E})$  and  ${\cal  P}_4  ({\cal  E})$,
respectively. As  also discussed above, ${\cal P}_4  ({\cal E})$ would
be universal for the two-qubit  density matrices and would involve the
Harr  measure on  $SU(4)$  \cite{tbs02}. This  is  depicted in  figure
(\ref{fig:p4}) and is common to all the two-qubit PDF of entanglement.

Now we  consider the  ${\cal P}_2 ({\cal  E})$ and ${\cal  P}_3 ({\cal
E})$ density functions for some representative states of the two qubit
system,  both for the  independent as  well as  collective decoherence
models. This enables us to  compare the entanglement in the respective
subspaces  of  the  system  Hilbert  space.  We  also  plot  the  full
entanglement density function curve ${\cal P}(\cal E)$ with respect to
the  entanglement ${\cal  E}$, at  a particular  time $t$.   This will
enable us  to look  at the contribution  to the entanglement  from the
different projections.

\begin{figure}
\subfigure[]{\includegraphics[width=7.0cm]{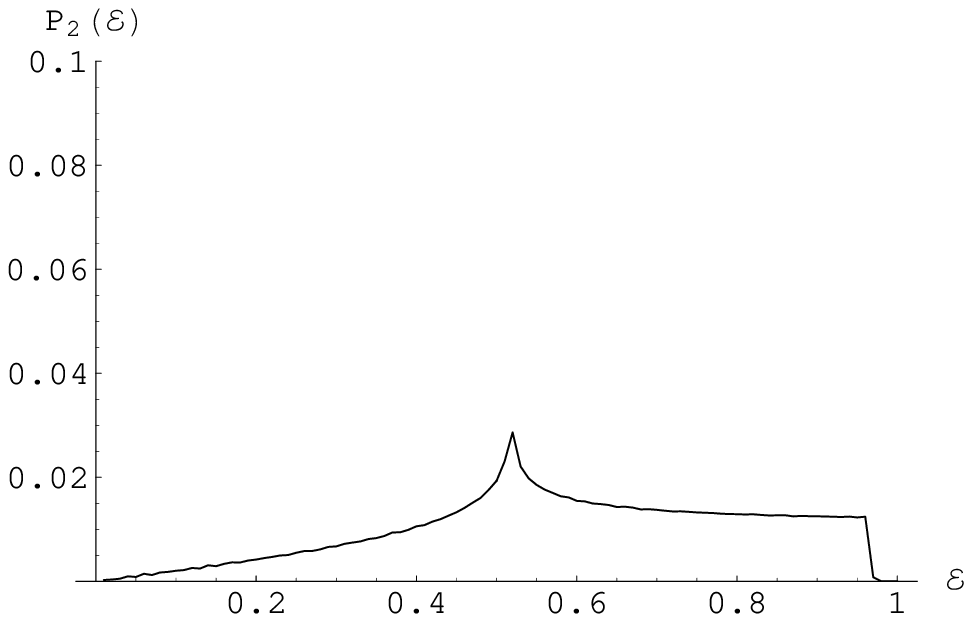}} 
\hfill
\subfigure[]{\includegraphics[width=7.0cm]{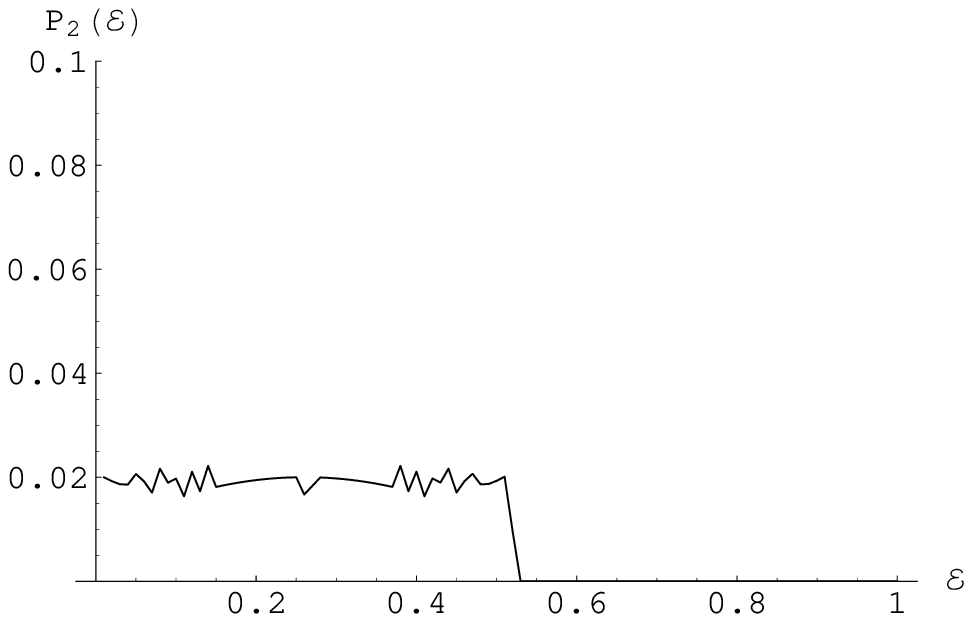}}
\caption{The density function ${\cal  P}_2 ({\cal E})$ with respect to
the entanglement ${\cal E}$ for an evolution time $t = 5.0$, $T = 0.0$
and bath squeezing  parameter $r$ equal to $0$.   Figure (a) refers to
the  independent   decoherence  model   and  (b)  to   the  collective
decoherence model.  }
\label{fig:P2t5}
\end{figure}

\begin{figure}
\subfigure[]{\includegraphics[width=7.0cm]{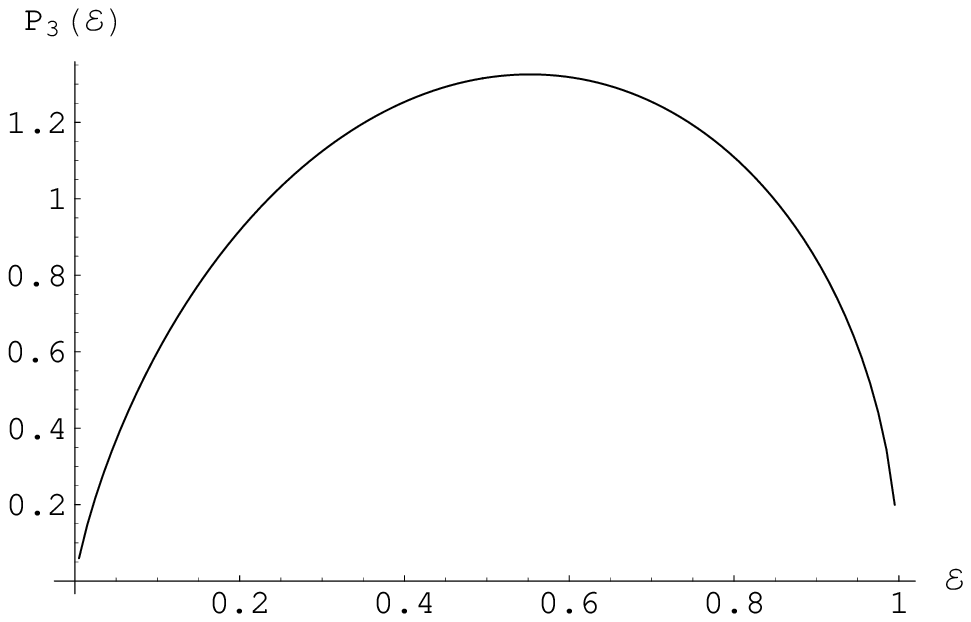}} 
\hfill
\subfigure[]{\includegraphics[width=7.0cm]{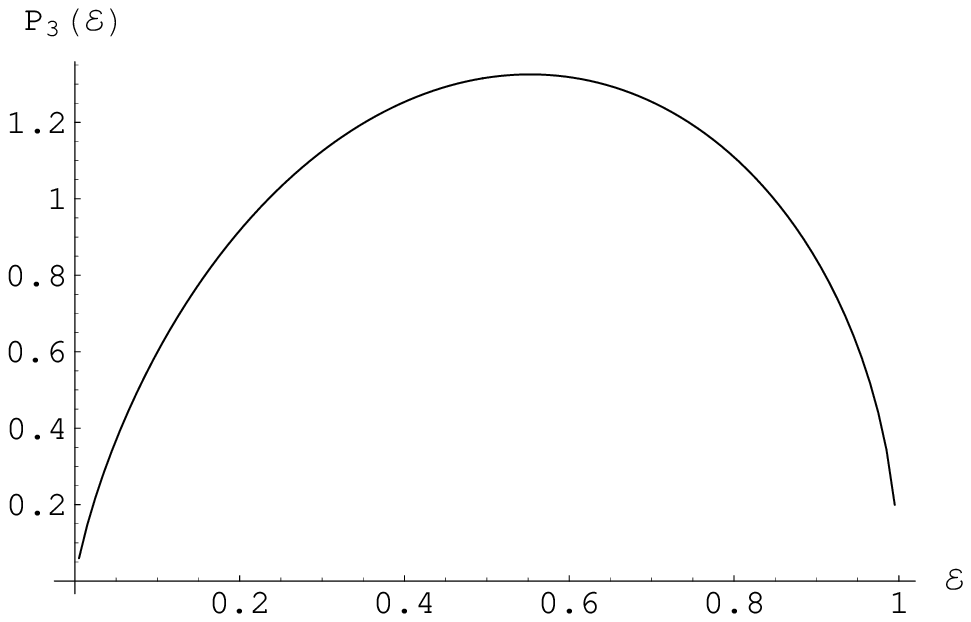}}
\caption{The density function ${\cal  P}_3 ({\cal E})$ with respect to
the entanglement ${\cal E}$ for an  evolution time $t = 5.0$, and zero
$T$  and  bath  squeezing.   Figure  (a)  refers  to  the  independent
decoherence model and (b) to the collective decoherence model.  }
\label{fig:P3t5}
\end{figure}

\begin{figure}
\subfigure[]{\includegraphics[width=7.0cm]{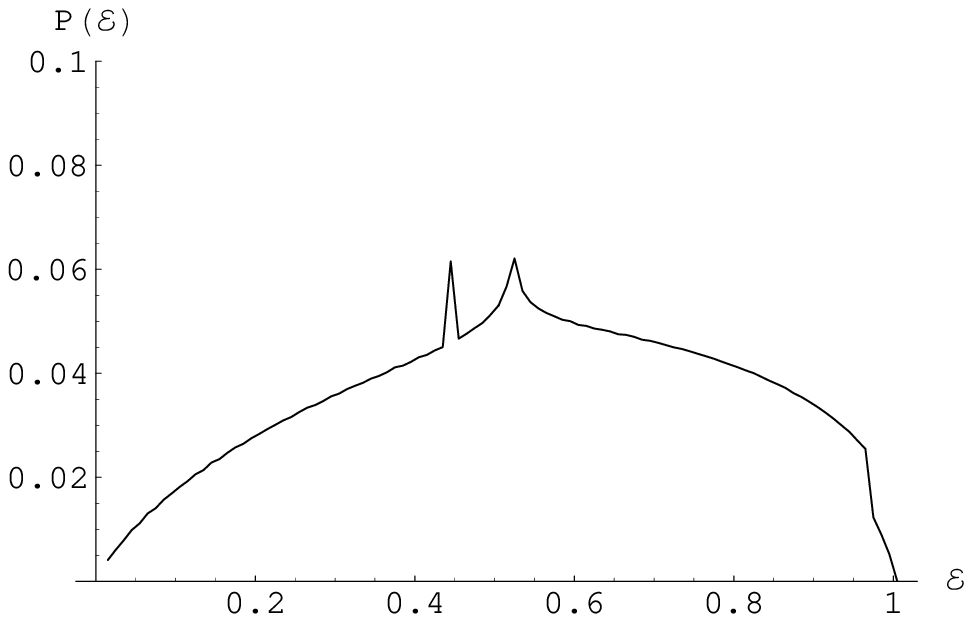}} 
\hfill
\subfigure[]{\includegraphics[width=7.0cm]{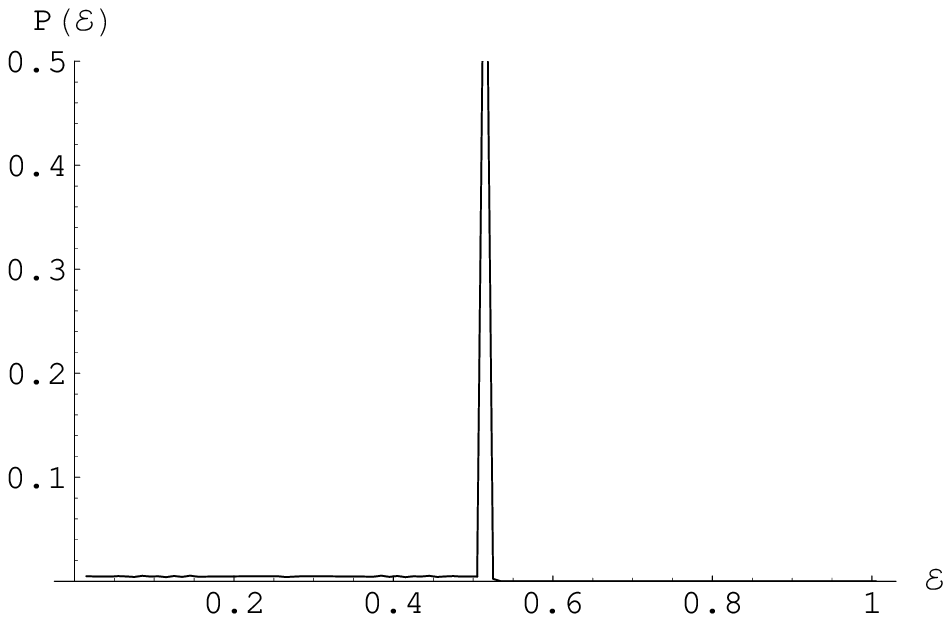}}
\caption{The  full density function  ${\cal P}({\cal  E})$ (\ref{PDF})
with respect to the entanglement ${\cal E}$ for an evolution time $t =
5.0$,  and zero  $T$ and  bath squeezing.   Figure (a)  refers  to the
independent decoherence  model and  (b) to the  collective decoherence
model.  }
\label{fig:Pfullt5}
\end{figure}

\begin{figure}
\subfigure[]{\includegraphics[width=7.0cm]{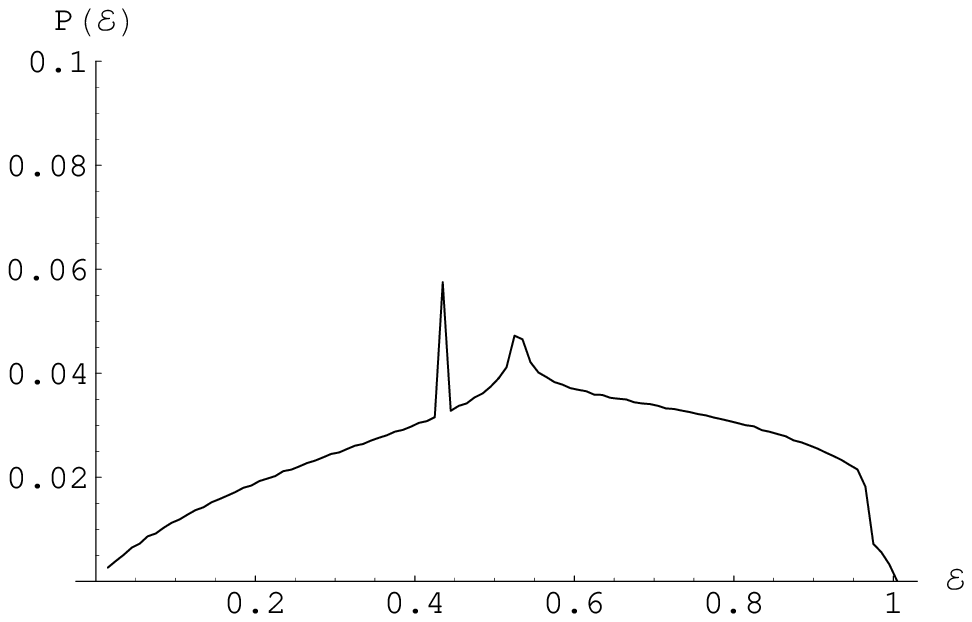}} 
\hfill
\subfigure[]{\includegraphics[width=7.0cm]{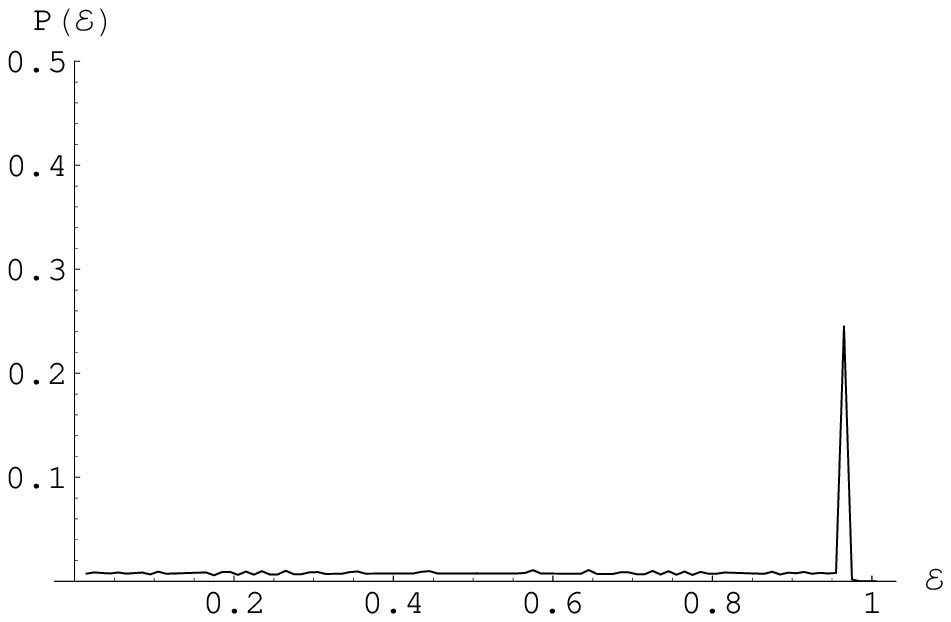}}
\caption{The  full density function  ${\cal P}({\cal  E})$ (\ref{PDF})
with respect to the entanglement ${\cal E}$ for an evolution time $t =
20.0$,  and zero $T$  and bath  squeezing.  Figure  (a) refers  to the
independent decoherence  model and  (b) to the  collective decoherence
model.  }
\label{fig:Pfullt20}
\end{figure}

Figures  (\ref{fig:P2t5}  (a)) and  (b)  depict  the  behavior of  the
density function ${\cal  P}_2 ({\cal E})$ for the  bath evolution time
$t = 5.0$  and $T = 0$ for the  independent and collective decoherence
models, respectively.  For these  conditions, the value of concurrence
(\ref{concur})  is 0.17 for  the case  of the  independent decoherence
model  and  0.42  for  the  collective model,  depicting  the  greater
entanglement content in the later compared to the former. This is also
borne out by these figures.   As shown in \cite{br08}, the concurrence
for  a  two  dimensional  projection  is ${\cal  C}_{\Pi_2}  =  ({\cal
  E}_{max} - {\cal  E}_{cusp})/2$.  From the figures (\ref{fig:P2t5}),
it can be seen that the value of ${\cal C}_{\Pi_2}$ is greater for the
case  of figure (\ref{fig:P2t5}(b))  when compared  to that  of figure
(\ref{fig:P2t5}(a)).  In  all the  figures related to  the probability
density functions of entanglement for the two-qubit system interacting
with an unsqueezed vacuum bath, for the independent decoherence model,
$kr_{12}$ is  set equal to  1.5, while for the  collective decoherence
model, $kr_{12}$ is  set equal to 0.07. For  the two-qubit interaction
with  a squeezed thermal  bath, the  independent decoherence  model is
parametrized  as above,  while for  the collective  decoherence model,
$kr_{12}$ is set equal to  0.08.  Figures (\ref{fig:P3t5} (a)) and (b)
exhibit  the entanglement  density  function ${\cal  P}_3 ({\cal  E})$
while  the  figures (\ref{fig:Pfullt5}  (a))  and  (b) illustrate  the
behavior of  the full density function  ${\cal P} ({\cal  E})$ for the
independent  and collective decoherence  models, respectively  and for
the same parameters as  above.  The full entanglement density function
is  obtained by  a  weighted sum  over  all the  contributions of  the
projection operators  from the different  subspaces (\ref{PDF}).  Here
and in all  the subsequent figures for the  full PDF, the contribution
from  the  one  dimensional  projection  $\Pi_1$,  which  is  a  delta
function,  is represented  by a  line of  height equal  to  its weight
(\ref{PDF})  and  the  point  on  the absicca  is  determined  by  its
corresponding entanglement.   The figures (\ref{fig:Pfullt5}  (a)) and
(b) depict  the rich entanglement  structure present in  the two-qubit
system  interacting   with  an  unsqueezed  vacuum   bath.  While  the
contribution   in  the   figure  (\ref{fig:Pfullt5}   (a)),   for  the
independent decoherence model, comes primarily from the two dimesional
projection, that  for the collective  decoherence model, as  in figure
(\ref{fig:Pfullt5}  (b)),  comes  from  the one  and  two  dimensional
projections  with the  one dimensional  projection being  the dominant
contributor. Since the one  dimensional projection represents the pure
state  entanglement  extant in  the  system,  this  clearly shows  the
greater entanglement  content in the collective  decoherence regime as
compared to the independent  one. The figures (\ref{fig:Pfullt20} (a))
and  (b) show  the behavior  of the  full density  function  ${\cal P}
({\cal  E})$ for  the independent  and collective  decoherence models,
respectively, for the bath evolution time $t = 20.0$ and $T = 0$.  For
these parameters, the value  of concurrence (\ref{concur}) is 0.32 for
the  case  of the  independent  decoherence  model  and 0.54  for  the
collective model,  depicting the  greater entanglement content  in the
later compared  to the former. By  a comparison with  the earlier case
for an evolution time $t = 5$, is brought out the point that with time
there  is a  build up  of entanglement  in the  two-qubit  system. The
system is still  possessed of a rich entanglement  structure.  For the
independent decoherence model,  the principal contribution still comes
from  the   two  dimensional   contribution  but  in   the  collective
decoherence  regime, in contrast  to the  earlier case,  the principal
contribution  has now  shifted from  the  one to  the two  dimensional
projection, thereby showing  that the system gets more  mixed with the
passage of time.

\begin{figure}
\subfigure[]{\includegraphics[width=7.0cm]{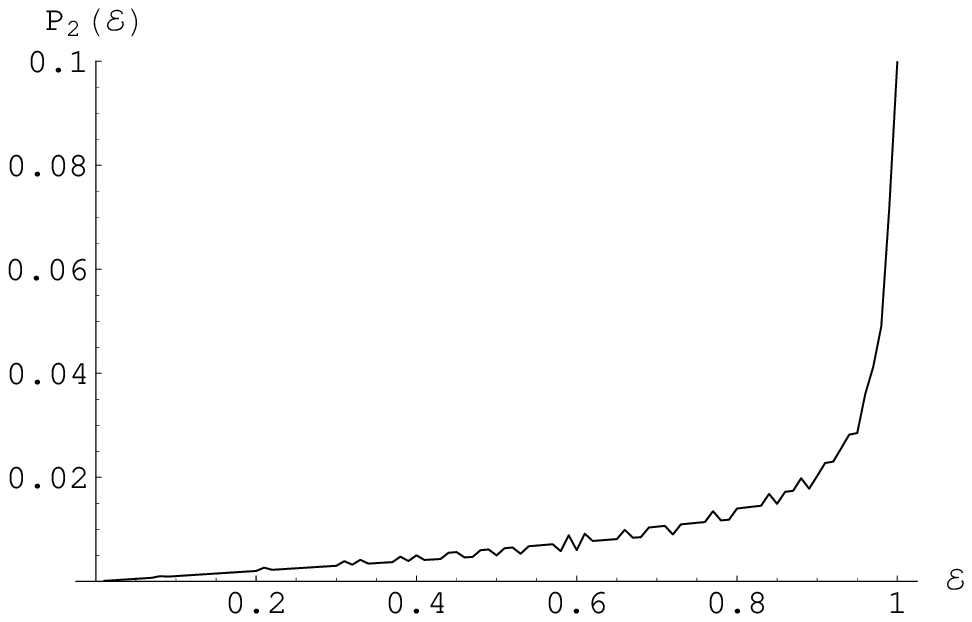}} 
\hfill
\subfigure[]{\includegraphics[width=7.0cm]{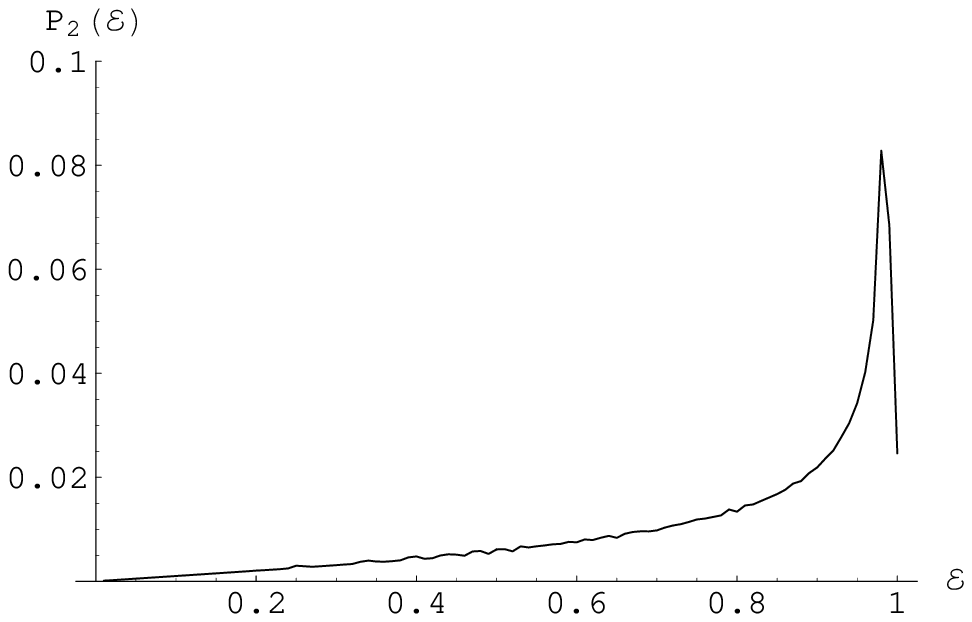}}
\caption{The density function ${\cal  P}_2 ({\cal E})$ with respect to
the entanglement  ${\cal E}$  for an  evolution time $t  = 1.0$,  $T =
10.0$ and  bath squeezing  parameter $r$ equal  to $0.5$.   Figure (a)
refers to the independent decoherence  model and (b) to the collective
decoherence model.  }
\label{fig:P2t1T10}
\end{figure}

\begin{figure}
\subfigure[]{\includegraphics[width=7.0cm]{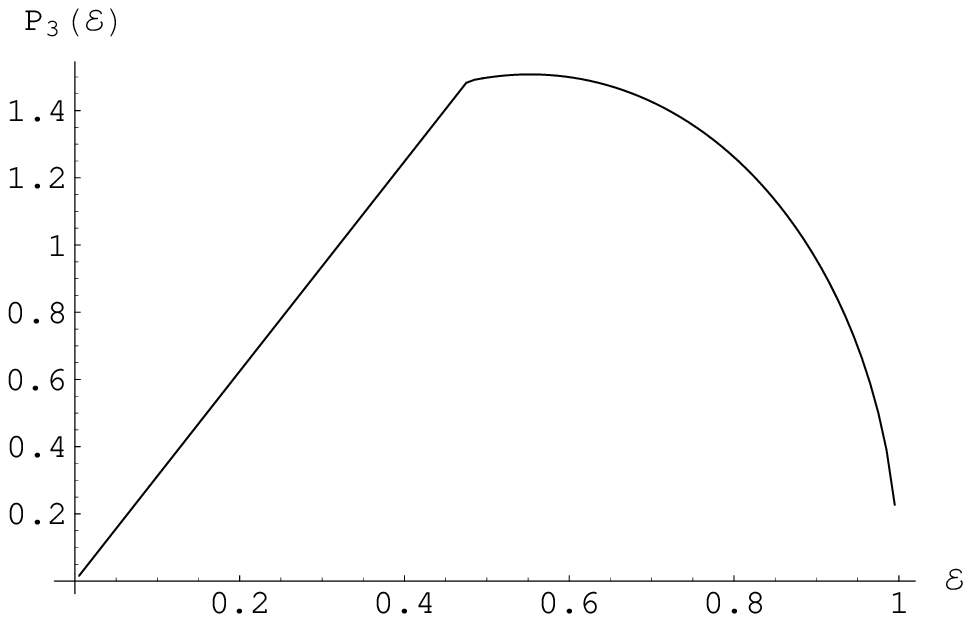}} 
\hfill
\subfigure[]{\includegraphics[width=7.0cm]{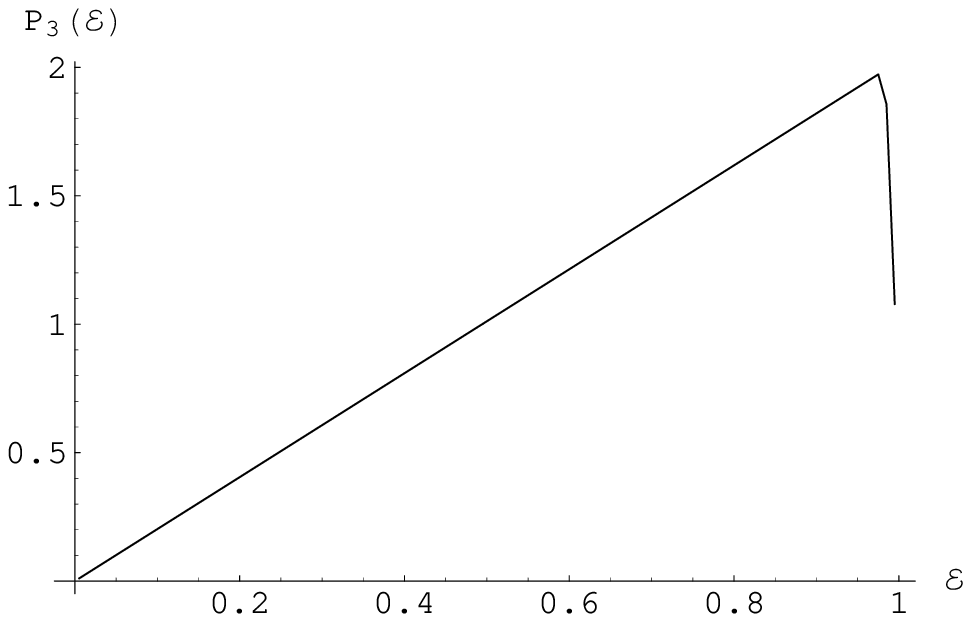}}
\caption{The density function ${\cal  P}_3 ({\cal E})$ with respect to
the entanglement  ${\cal E}$  for an  evolution time $t  = 1.0$,  $T =
10.0$ and  bath squeezing  parameter $r$ equal  to $0.5$.   Figure (a)
refers to the independent decoherence  model and (b) to the collective
decoherence model.  }
\label{fig:P3t1T10}
\end{figure}

\begin{figure}
\subfigure[]{\includegraphics[width=7.0cm]{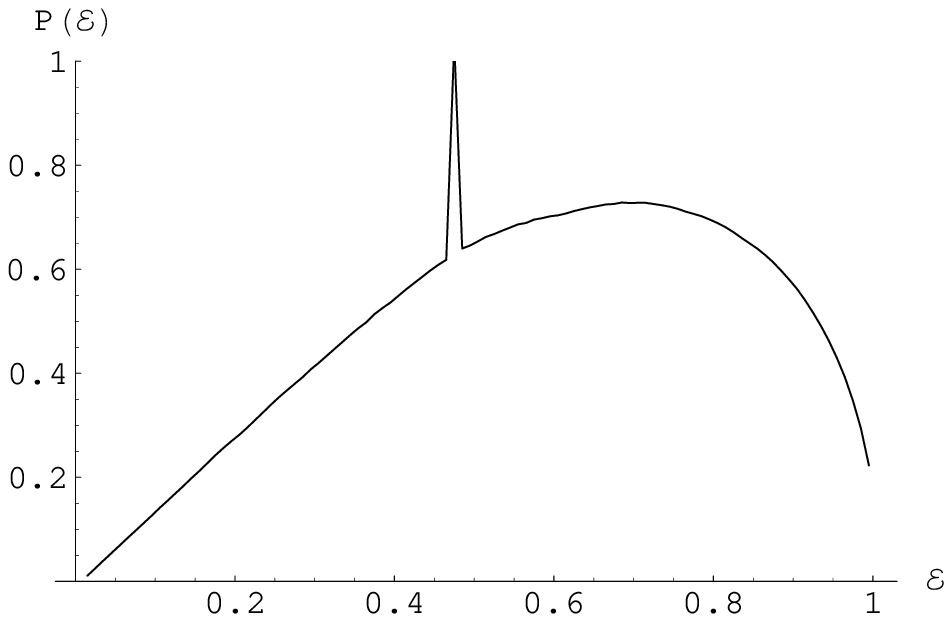}} 
\hfill
\subfigure[]{\includegraphics[width=7.0cm]{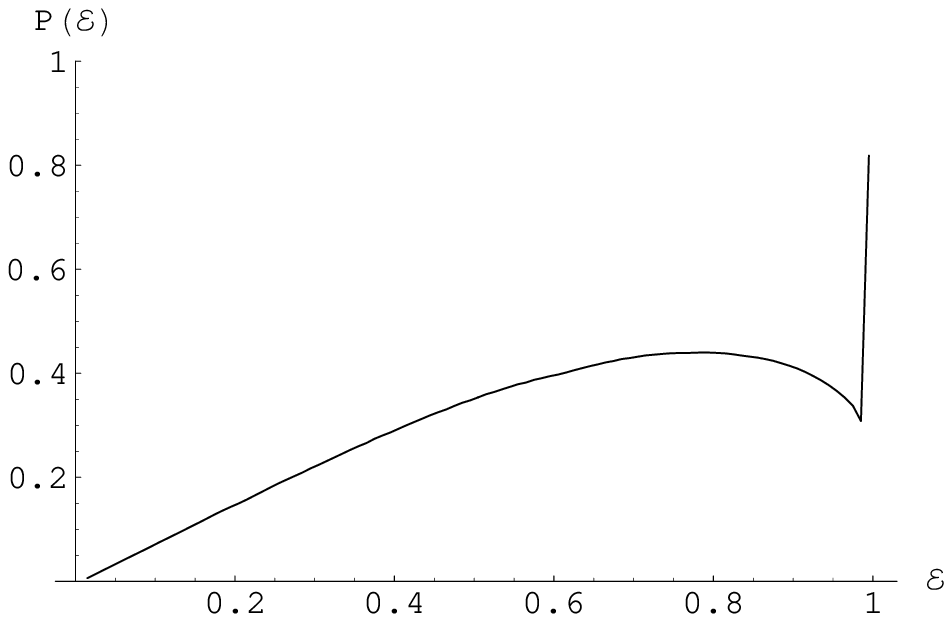}}
\caption{The  full density function  ${\cal P}({\cal  E})$ (\ref{PDF})
  with respect to the entanglement ${\cal E}$ for an evolution time $t
  = 1$,  $T = 10.0$ and  bath squeezing parameter $r$  equal to $0.5$.
  Figure (a)  refers to the  independent decoherence model and  (b) to
  the collective decoherence model.}
\label{fig:Pfullt1T10r0p5}
\end{figure}

We now  consider the  nature of entanglement  in the  two-qubit system
when   it   interacts  with   a   squeezed   thermal  bath.    Figures
(\ref{fig:P2t1T10}   (a))  and   (b)  depict   the  behavior   of  the
entanglement  density  function  ${\cal   P}_2  ({\cal  E})$  for  the
evolution  time  $t  = 1$,  $T  =  10$  and bath  squeezing  parameter
(\ref{N},  \ref{M})  $r =  0.5$  for  the  independent and  collective
decoherence models, respectively.  For  these conditions, the value of
concurrence (\ref{concur})  is 0,  indicating a complete  depletion of
entanglement.  This is  partially borne out by the  fact that for this
case ${\cal C}_{\Pi_2} \approx 0$.   However, as seen from the figures
(\ref{fig:P3t1T10}   (a))  and   (b)   and  also   from  the   figures
(\ref{fig:Pfullt1T10r0p5} (a))  and (b),  the system still  exhibits a
rich  entanglement structure. In  figure (\ref{fig:P3t1T10}  (a)), the
parameter ${\cal  E}_{\perp}$ is $\approx  0.48$ while it  is $\approx
0.98$ for  the collective model as in  (\ref{fig:P3t1T10} (b)).  Since
${\cal E}_{\perp}$  is evaluated from  the state perpendicular  to the
third  non-separable   basis  in  the  canonical  basis   of  a  three
dimensional  projection  \cite{br08},  this  clearly  brings  out  the
greater entanglement  content in the  collective model as  compared to
the independent  case.  In figure  (\ref{fig:Pfullt1T10r0p5} (a)), the
contribution  to the  full  entanglement density  for the  independent
model comes from the one, two and four dimensional projections with an
approximately  equal contribution  from the  one and  four dimensional
projections  while  in   figure  (\ref{fig:Pfullt1T10r0p5}  (b)),  the
contribution to the full entanglement density for the collective model
comes  from  the  one   and  four  dimensional  projections  with  the
contribution from  the one  dimensional projection being  the dominant
one. Since the contribution from  the one dimensional projection is an
indicator  of pure  state  entanglement in  the  system, this  clearly
brings out  the entanglement  content in the  system, under  the given
conditions, even  though concurrence is zero. Also  the greater weight
carried by the  one dimensional projection in the  collective model in
comparison  with   the  independent   one,  brings  out   the  greater
entanglement content in the former.

\begin{figure}
\subfigure[]{\includegraphics[width=7.0cm]{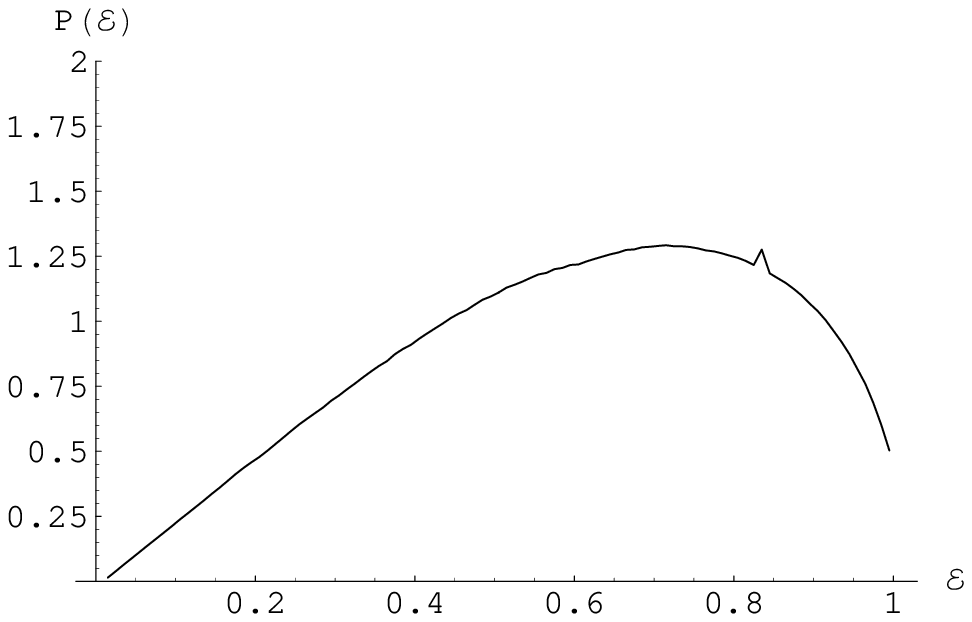}} 
\hfill
\subfigure[]{\includegraphics[width=7.0cm]{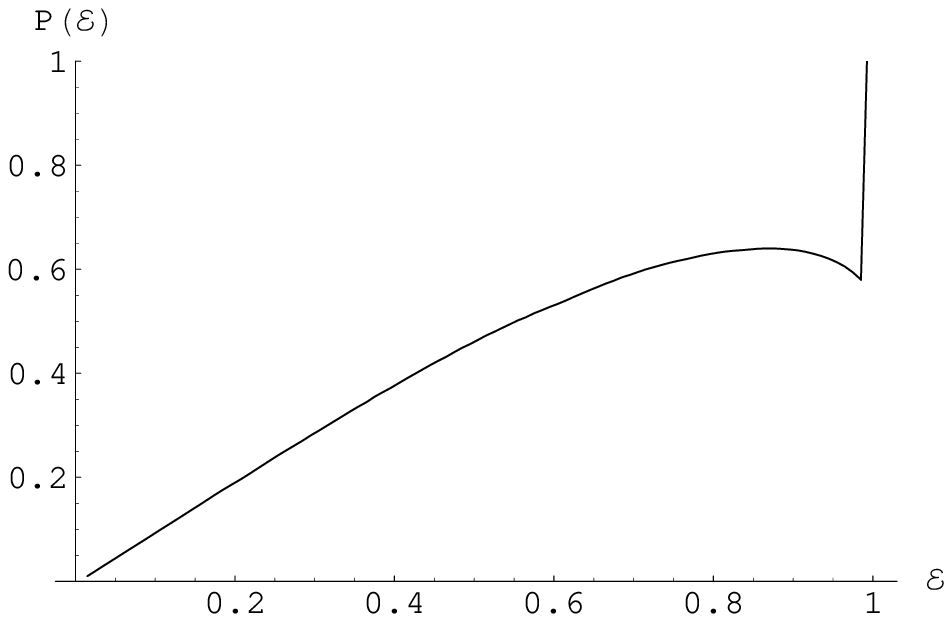}}
\caption{The  full density function  ${\cal P}({\cal  E})$ (\ref{PDF})
  with respect to the entanglement ${\cal E}$ for an evolution time $t
  = 1$,  $T = 10.0$ and  bath squeezing parameter $r$  equal to $1.0$.
  Figure (a)  refers to the independent decoherence  model and (b)  to the  
collective  decoherence model.}
\label{fig:Pfullt1T10r1p0}
\end{figure}

\begin{figure}
\subfigure[]{\includegraphics[width=7.0cm]{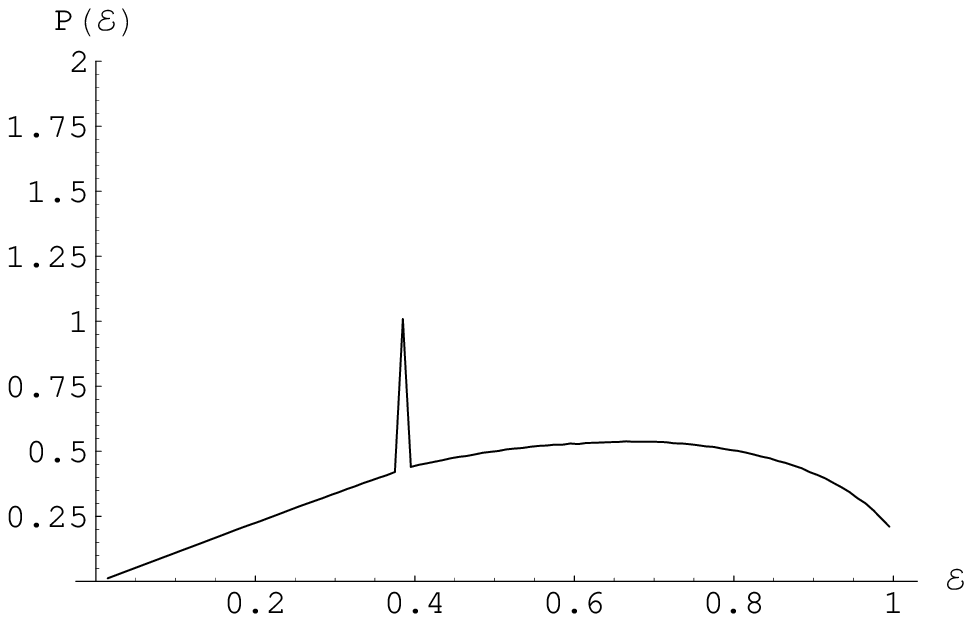}} 
\hfill
\subfigure[]{\includegraphics[width=7.0cm]{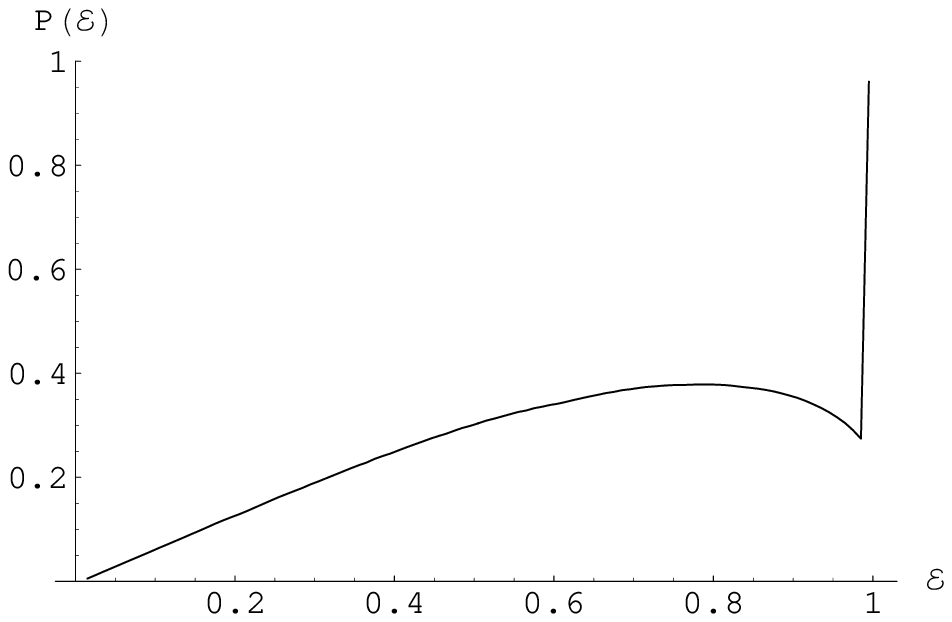}}
\caption{The  full density function  ${\cal P}({\cal  E})$ (\ref{PDF})
  with respect to the entanglement ${\cal E}$ for an evolution time $t
  = 1$, $T = 10.0$ and bath squeezing parameter $r$ equal to $0$, i.e,
  we consider here  an unsqueezed thermal bath.  Figure  (a) refers to
  the  independent  decoherence  model   and  (b)  to  the  collective
  decoherence model.}
\label{fig:Pfullt1T10r0p0}
\end{figure}

\begin{figure}
\subfigure[]{\includegraphics[width=7.0cm]{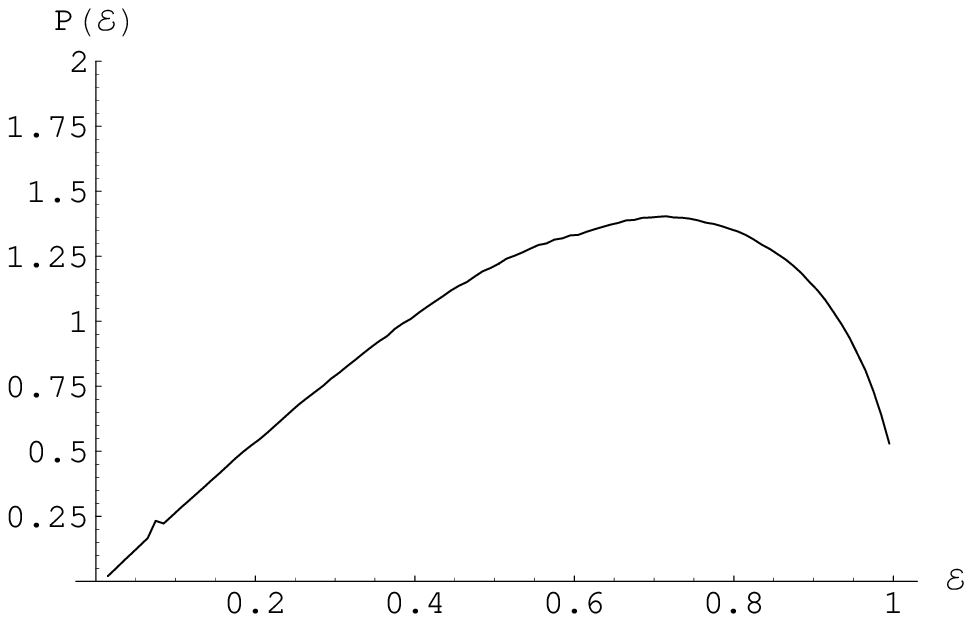}}
\hfill
\subfigure[]{\includegraphics[width=7.0cm]{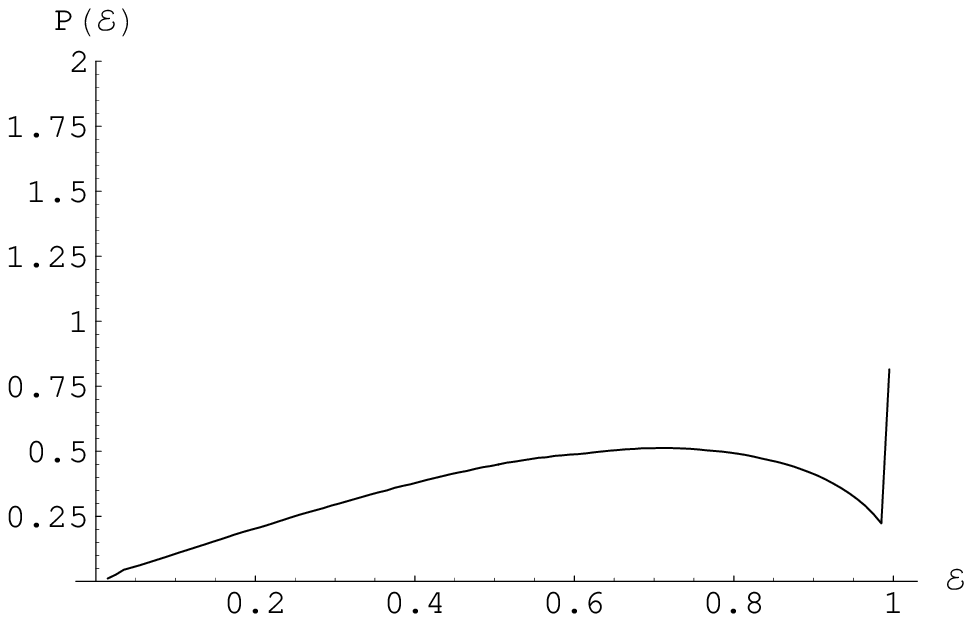}}
\caption{The  full density function  ${\cal P}({\cal  E})$ (\ref{PDF})
  with respect to the entanglement ${\cal E}$ for an evolution time $t
  = 5$,  $T = 10.0$ and  bath squeezing parameter $r$  equal to $0.5$.
  Figure (a)  refers to the  independent decoherence model and  (b) to
  the collective decoherence model.}
\label{fig:Pfullt5T10}
\end{figure}

Figures  (\ref{fig:Pfullt1T10r1p0} (a))  and (b)  illustrate  the full
entanglement density  for the evolution  time $t =  1$, $T =  10$ and
bath  squeezing  parameter  (\ref{N},  \ref{M})  $r  =  1.0$  for  the
independent and collective decoherence models, respectively. For these
conditions, the value of  concurrence (\ref{concur}) is zero. However,
the system still exhibits an entanglement structure, especially in the
collective    decoherence    regime   as    seen    in   the    figure
(\ref{fig:Pfullt1T10r1p0}   (b))  where  the   one,  three   and  four
dimensional  projections contribute  with  the principal  contribution
coming  from the  one dimensional  projection, thereby  indicating the
presence of  pure state entanglement in the  system. The corresponding
case for the independent model, as in figure (\ref{fig:Pfullt1T10r1p0}
(a)), has its principal  contribution coming from the four dimensional
projection representing  a maximally  mixed state. This  clearly shows
the richer entanglement content  in the collective model in comparison
to the independent one. Figures (\ref{fig:Pfullt1T10r0p0} (a)) and (b)
illustrate the  full entanglement density, for the  same conditions as
above but with zero bath squeezing, for the independent and collective
decoherence models,  respectively. The  entanglement in the  system is
seen to be  greater for this case when compared  to the previous cases
with finite bath squeezing. The value of concurrence (\ref{concur}) is
zero for  the independent model, while  it is 0.17  for the collective
one, bringing  out the  greater entanglement content  in the  later in
comparison with the former. This is also borne out by the entanglement
density function, where the  contribution to the full density function
from  the  one  dimensional   projection  (indicative  of  pure  state
entanglement) is greater in the  collective model when compared to the
independent one. Also, in this  case of interaction with an unsqueezed
thermal  bath,   the  full  entanglement  density   function  for  the
independent  decoherence  model  exhibits  greater  entanglement  when
compared  to the  corresponding case  of interaction  with  a squeezed
thermal bath. This seems to  indicate that for the two-qubit system, a
finite   bath  squeezing   is  detrimental   to  the   development  of
entanglement. Finally in figures (\ref{fig:Pfullt5T10} (a)) and (b) is
depicted the full entanglement density  for the evolution time $t = 5$
, $T =  10$ and bath squeezing parameter (\ref{N},  \ref{M}) $r = 0.5$
for    the   independent    and    collective   decoherence    models,
respectively. As  expected, with the  increase in the exposure  to the
environment,  indicated  by the  greater  evolution  time, the  system
becomes more mixed and hence loses entanglement. This is borne out by
the  fact  that  for   these  conditions,  the  value  of  concurrence
(\ref{concur}) is zero. However, for the collective decoherence model,
as  shown  in  figure  (\ref{fig:Pfullt5T10} (b)),  the  full  density
function has  contributions coming from  the one and  four dimensional
projections, with the weight carried by the one dimensional projection
greater than  that by the four  dimensional one bringing  out the fact
that pure state  entanglement is still extant in  the system under the
given conditions.  However, for  the independent decoherence model, as
in  figure (\ref{fig:Pfullt5T10} (a)),  the principal  contribution to
the full entanglement density function comes from the four dimensional
projection, indicating that the  system is tending towards a maximally
mixed state, thereby losing its entanglement.

Thus we see that when compared to  the interaction with a bath at $T =
0$,  the  finite  $T$  bath   is  detrimental  to  the  generation  of
entanglement between the  two-qubit system.  An interesting connection
of this can be made with  the work presented in \cite{dn04}, where the
authors   connected   frustration   with  interaction   strength   and
ground-state   entanglement.    They   defined  frustration   as   the
non-commutativity of  the local Hamiltonian (consisting  of the single
body terms) with the interaction  Hamiltonian and argued that with the
increase in interaction, ground-state entanglement in the system would
increase.   The  case of  dissipative  $S-R$  interaction with  $[H_S,
H_{SR}]   \neq  0$  would   suggest  a   frustrated  system   in  this
nomenclature.   For the  interaction of  the two-qubit  system  with a
vacuum  bath,  entanglement is  seen  to  rise  with increase  in  the
influence  of the  environment qualitatively  agreeing with  the above
work.   Finally  entanglement would  vanish  because  as  a result  of
decoherence,  the system would  lose its  quantum coherence  and tend
towards the classical regime.

\section{An application to quantum communication: quantum 
repeaters}
The technique of entanglement purification \cite{chb96} can be adapted
for quantum communication  over long distances, the key  idea behind a
{\it  quantum  repeater}  \cite{bdcz98}.   The efficiency  of  quantum
communication  over long  distances is  reduced due  to the  effect of
noise, which can  be considered as a natural  open system effect.  For
distances much  longer than  the coherence length  of a  noisy quantum
channel, the fidelity of transmission  is usually so low that standard
purification  methods  are  not  applicable.  In  a  quantum  repeater
set-up, the channel is divided into shorter segments that are purified
separately  and   then  connected   by  the  method   of  entanglement
swapping. This method can be much more efficient than schemes based on
quantum  error  correction,  as  it  makes  explicit  use  of  two-way
classical   communication.   The   quantum   repeater  system   allows
entanglement purification  over arbitrary long  channels and tolerates
errors  on the  percent level.  It requires  a polynomial  overhead in
time,   and  an   overhead  in   local  resources   that   grows  only
logarithmically with the length of the channel.

We  consider  the  effect  of  noise, introduced  by  imperfect  local
operations that constitute the  protocols of entanglement swapping and
purification ,  on such  a compound  channel, and how  it can  be kept
below  a certain  threshold.  The  noise  process studied  is the  one
obtained  from  the  two-qubit  reduced  dynamics  via  a  dissipative
system-reservoir   interaction,   studied   above,  instead   of   the
depolarizing noise  considered in  \cite{bdcz98}.  Here we  treat this
problem  in a  simplified  fashion, and  study  the applicability  and
efficiency of entanglement purification  protocols in the situation of
imperfect local  operations. A  treatment of partial  teleportation of
entanglement, conceptually equivalent to  quantum entanglement
swapping, in a noisy environment was made in \cite{lkpl}.

A quantum  repeater involves the  two tasks of  entanglement swapping,
involving  Bell-state  measurements,  and  entanglement  purification,
involving CNOT  gates. The Bell-state measurement  may be equivalently
replaced   by   a  CNOT   followed   by   a  projective   single-qubit
measurement. In  entanglement swapping, two  distant parties initially
not  sharing entanglement  with each  other, but  sharing entanglement
separately  with  a third  party,  become  entangled  by virtue  of  a
multi-partite  measurement by  the  third party  on  the latter's  two
halves  of  entanglement.    Entanglement  purification  involves  two
parties employing local  operations and classical communication (LOCC)
to  improve the  fidelity $F$  of Einstein-Podolsky-Rosen  (EPR) pairs
they share,  with respect to  a maximally entangled state.   The local
operations  involve  two-qubit  gates  such  as  the  CNOT  operation,
followed by single qubit measurement,  and a possible discarding of an
EPR pair. Provided $F>0.5$, and  at the cost of losing shared (impure)
entanglement,  the  two  parties  can  increase the  fidelity  of  the
remaining shared entanglement to
\begin{equation}
F^\prime = \frac{F^2 + [(1-F)/3]^2}{F^2 + [2F(1-F)/3] + (5/9)(1-F)^2},
\label{eq:bennett}
\end{equation}
where  $F$ and  $F^\prime$  are, respectively,  the  input and  output
fidelities  of  the  entanglement  purification protocol  proposed  by
Bennett {\it et al.} \cite{chb96}.

In the  simplified scenario considered  here, the output of  the noisy
CNOT  is taken  to be  a mixed  separable state,  in place  of  a pure
separable state that is obtained  in the noiseless case.  As a further
simplification,  in order  to faciliate  an easy  connection  with the
purification  protocol due  to Bennett  et  al., this  mixed state  is
assumed to be of the form:
\begin{equation}
\rho(F) = F^2|\uparrow,\downarrow\rangle\langle \uparrow,\downarrow| + 
(1-F^2)|\downarrow,\uparrow\rangle\langle \downarrow,\uparrow|,
\label{eq:inrho}
\end{equation}
where $|\uparrow\rangle~~ (|\downarrow\rangle) \equiv 
|\frac{1}{2}\rangle~~ (|-\frac{1}{2}\rangle)$.
Thus, $\rho(F)$ is  a mixture in the two  dimensional space spanned by
$\{|\uparrow,\downarrow\rangle,         |\downarrow,\uparrow\rangle\}$,
parametrized    by    fidelity    $F$,   given    by    $\sqrt{\langle
\uparrow,\downarrow|\rho(F)|\uparrow,\downarrow\rangle}$.   The  state
$\rho(F)$ is then  the input to the purification  protocol, whereby we
obtain the output fidelity $F^\prime$ as a function of $F$.

\begin{figure}
\subfigure[]{\includegraphics[width=7.0cm]{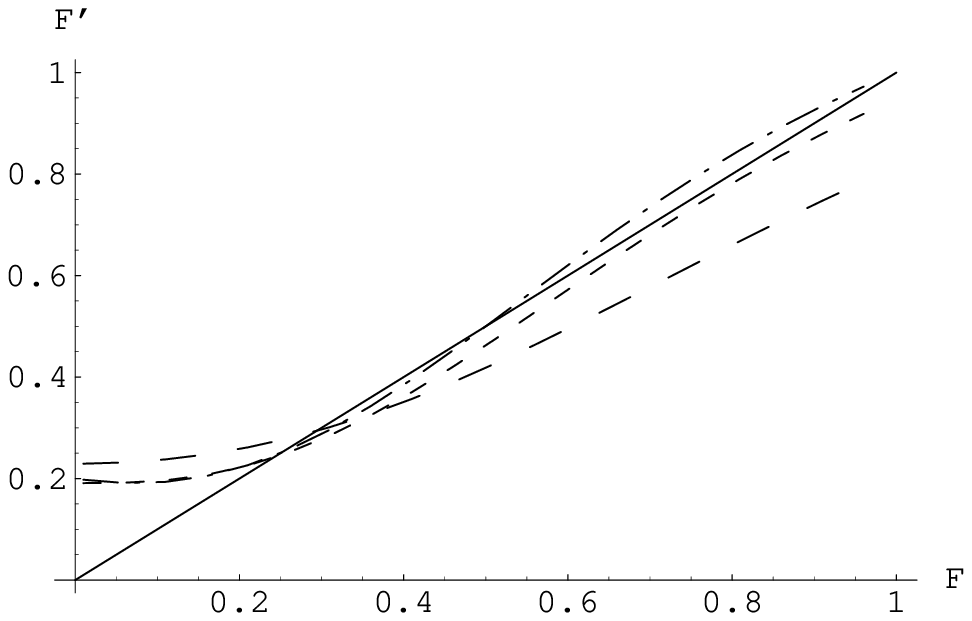}}
\hfill
\subfigure[]{\includegraphics[width=7.0cm]{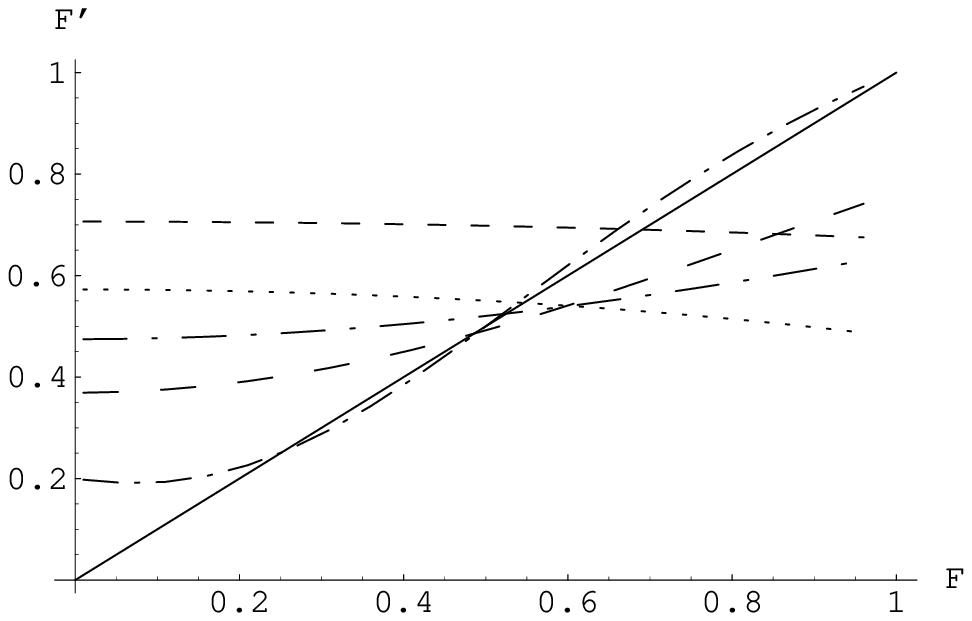}}
\caption{Purification loop for connecting and purifying EPR pairs. The
  noisy  channel is  modelled as  a two-qubit  dissipative interaction
  with an unsqueezed vacuum bath in the independent $(kr_{ab} \geq 1)$
  (plot (a))  and collective $(kr_{ab} \ll 1)$  (plot (b)) decoherence
  regime,   with    the   input   state   given    by   $\rho(F)$   in
  Eq. (\ref{eq:inrho}).   The bold line is the  $F=F^\prime$ plot, the
  small  and   large  dashed   curves  represent  $t=3$   and  $t=10$,
  respectively,  while the dot-dashed  curve is  due to  the noiseless
  Bennett et al. protocol.  In  (b), the larger dash-dot curve and the
  dotted   curves   represent,   respectively,  $t=14$   and   $t=20$,
  respectively.}
\label{fig:rep2QVacindep1collc} 
\end{figure}



We depict  in figures (\ref{fig:rep2QVacindep1collc} (a))  and (b) the
modified  `purification loop',  obtained by  subjecting  the noiseless
loop to  the above model of  noise for the  independent and collective
decoherence regimes, respectively.  In both figures we find that noise
degrades the  performance of  the purification protocol.   To evaluate
its performance, it may be  compared with the noiseless case, given by
the dot-dashed curve  in both figures. That this  curve lies above the
$F=F^\prime$  line   in  the  closed  range   $[0.5,1]$  implies  that
fidelities above the minimum  value $F_{\rm min}=0.5$ can be corrected
to  the  maximum value  $F_{\rm  max}=1$  by  repeated application  of
purification.  The degrading effect of  noise can be seen in two ways:
it introduces an off-set, whereby an input of $F=1$ does not yield the
same output; it restricts $F_{\rm max}$ to values less than 1.

In figure (\ref{fig:rep2QVacindep1collc} (a)), comparison of the small
and  large dashed  curves  shows that  increasing  bath exposure  time
degrades the fidelity by suppressing  $F_{\rm max}$ (the point where a
curve  cuts  the $F=F^\prime$  line  from  above)  and increasing  the
off-set at $F=1$.  Although figure (\ref{fig:rep2QVacindep1collc} (b))
similarly  shows the  expected off-set  due to  noise,  the surprising
feature is  that the curves  for $t=3$ and  $t=20$ show a  slight {\it
  lowering} of  output fidelity  with increasing input  fidelity.  The
matter is compounded by noting  that at an intermediate time ($t=14$),
as depicted by the large-dash-dot curve, the function $F^\prime(F)$ is
monotonously  {\it increasing}.  This  may be  attributed to  the fact
that entanglement generated  due to interaction with the  bath shows a
strong oscillatory behavior,  as seen from figure (\ref{fig:concuTim})
for the collective regime.

\section{Conclusions}

Here  we have  analyzed the  dynamics of  entanglement in  a two-qubit
system  interacting with  its environment,  taken to  be in  a general
squeezed  thermal  state, via  a  dissipative  $S-R$ interaction.  The
analysis of the mixed state entanglement has been made using a measure
involving a probability density function (PDF).

The position dependent coupling of  the qubits with the bath enabled a
natural division  of the dynamics  into an independent  and collective
decoherence regime,  where in  the independent decoherence  regime the
qubits  interact  via  localized  $S-R$  interactions,  while  in  the
collective  regime  the qubits  are  close  enough  to feel  the  bath
collectively.   The reduced  dynamics revealed  that in  this  case of
dissipative $S-R$  interaction, there is  no decoherence-free subspace
in contrast to that of  a QND $S-R$ interaction \cite{brs1}.  This can
be understood in  the following manner.  As shown  in \cite{brs1}, the
two-qubit  QND reduced  dynamics  obeys the  relation $\rho_{ab}(t)  =
{\cal  L}_{ab}(t) \rho_{ab}(0)$,  with the  non-trivial aspect  of the
dynamics  being  that ${\cal  L}$  represents,  not  a matrix,  but  a
two-dimensional  array,  with  the multiplication  done  element-wise.
This gets translated into the spin-flip symmetry obeyed by the reduced
dynamics.   The  emergence of  a  decoherence-free  subspace could  be
attributed  to the  non-trivial  symmetry obeyed  by  the system.   An
absence  of  a  decoherence-free  subspace  in  the  present  case  of
dissipative interaction could be due to the lack of such a symmetry in
the reduced dynamics.

We studied the entanglement in the two-qubit system for different bath
parameters,  for both  vacuum as  well  as a  bath at  finite $T$  and
squeezing and used the measure of mixed state entanglement involving a
PDF  as well  as concurrence.   It clearly  emerged  that entanglement
generation is more efficient in the case of interaction with a bath at
zero $T$  than at  finite $T$. It  also appears  that in this  case of
two-qubit  dissipative  $S-R$  interaction,  presence of  finite  bath
squeezing  does  not help  in  entanglement  generation.   This is  in
contrast to expectations from the single-qubit dissipative interaction
\cite{bsdiss} where it was observed that for some bath parameters, the
process  of  decoherence  could   be  slowed  down  resulting  in  the
preservation  of quantum  coherence for  a longer  time.  It  would be
pertinent to point out here  an interesting work \cite{pr08} where the
authors studied  in detail the  evolution of entanglement  between two
oscillators  coupled to  the same  environment.  There  the oscillator
system was  assumed to start  from a general two-mode  Gaussian state,
which includes the two-mode  squeezed state. Entanglement in the final
state was considered  as a resource whose origin  could be the quantum
resource  available in  the  system, viz.   squeezing  in the  initial
oscillator system.   It was  shown that under  certain regimes  of the
squeezing parameter,  the environment could  be used as a  resource to
extract entanglement, thereby highlighting the importance of squeezing
in  the study  of entanglement  generation.  The  work  presented here
differs from that in \cite{pr08}, apart from the obvious difference in
the system (here a two-qubit system), in that the evolution considered
here  is  Markovian while  that  in  \cite{pr08}  is that  of  quantum
Brownian  motion which  is essentially  non-Markovian. Also,  here the
squeezing originated  from the bath  while that in \cite{pr08}  was in
the initial system  state. It would be of interest  to extend the work
presented  here  to have  a  better  understanding  of the  impact  of
squeezing,   a  natural   resource,  on   the  dynamic   evolution  of
entanglement.

The  entanglement  analysis  via  the  PDF showed  an  advantage  over
concurrence, specially in the case  of mixed state entanglement due to
interaction with  a finite $T$  bath.  While concurrence was  zero for
most  of the  cases  considered,  the PDF  revealed  the structure  of
entanglement still  present in the  two-qubit system bringing  out the
statistical and geometrical nature of the measure.  Finally we made an
application  of  the  dissipative  two-qubit  reduced  dynamics  to  a
simplified  model of  a quantum  repeater,  which can  be adapted  for
quantum communication  over long distances. Thus this  work along with
\cite{brs1}  makes  a detailed  study  of  the  dynamics of  two-qubit
entanglement in the  presence of generic, purely dephasing  as well as
dissipative, open quantum system effects.

\acknowledgments

We wish to thank Shanthanu Bhardwaj for numerical help.


\begin{thebibliography}{100}

\bibitem{wl73} W. H. Louisell, {\it Quantum Statistical 
Properties of Radiation} (John Wiley and Sons, 1973). 

\bibitem{cl83} A. O. Caldeira and A. J. Leggett, Physica A {\bf 
121}, 587 (1983). 

\bibitem{wz93} W. H. Zurek, Phys. Today {\bf 44}, 36 (1991); 
Prog. Theor. Phys. {\bf 87}, 281 (1993).

\bibitem{bg07} S. Banerjee and R. Ghosh, J. Phys.
A: Math. Theo. {\bf 40}, 13735 (2007); eprint quant-ph/0703054.

\bibitem{ha85} V. Hakim and V. Ambegaokar, Phys. Rev. A {\bf 
32}, 423 (1985). 

\bibitem{sc87} C. M. Smith and A. O. Caldeira, Phys. Rev. A 
{\bf 36}, 3509 (1987); {\it ibid} {\bf 41}, 3103 (1990). 

\bibitem{gsi88} H. Grabert, P. Schramm and G. L. Ingold, 
Phys. Rep. {\bf 168}, 115 (1988). 

\bibitem{sb00} S. Banerjee and R. Ghosh, Phys. Rev. A {\bf 62}, 
042105 (2000). 

\bibitem{sb03-2} S. Banerjee and R. Ghosh, Phys. Rev. E {\bf 
67}, 056120 (2003). 

\bibitem{myatt}   C.  J.   Myatt,  B.   E.  King,   Q.   A.  Turchette,
C. A. Sackett, {\it et al.}, Nature  {\bf 403}, 269 (2000).

\bibitem{turch} Q. A. Turchette,
C. J. Myatt, B. E. King, C. A. Sackett, {\it et al.}, Phys. Rev. A {\bf 62},
053807 (2000).

\bibitem{bell} J. S. Bell, Physics {\bf 1}, 195 (1964). 

\bibitem{nc} M. Nielsen and I. Chuang, {\it Quantum Computation
and Quantum Information} (Cambridge University Press, Cambridge, 2000).

\bibitem{shor} P. Shor, SIAM Journal of Computing {\bf 26}, 1484 (1997);
L. K. Grover, Phys. Rev. Lett. 79, 325 (1997).

\bibitem{css} S. Calderbank and P. Shor, Phys. Rev. {\b A 54}, 1098 (1996);
A. Steane, Proc. Roy. Soc., London, Ser. A {\bf 452}, 2551 (1996).

\bibitem{beige} A. Beige, D. Braun, B. Tregenna and P. L. Knight, 
Phys. Rev. Lett. {\bf 85}, 1762 (2000). 

\bibitem{fy00} M. Fleischhauer, S. F. Yelin and M. D. Lukin, 
Opt. Commun. {\bf 179}, 395 (2000). 

\bibitem{sm02} S. Schneider and G. J. Milburn, Phys. Rev. A {\bf 65}, 
042107 (2002). 

\bibitem{bp02} H.-P. Breuer and F. Petruccione, \textit{The Theory 
of Open Quantum Systems} (Oxford University Press 2002).

\bibitem{wlk8} D. Wilson, J. Lee and M. S. Kim, 
J. Mod. Optics {\bf 50}, 1809 (2003).

\bibitem{simon} R. Simon, Phys. Rev. Lett. {\bf 84}, 2726 (2000).

\bibitem{pa08} L. D. Contreras-Pulido and R. Aguado, 
Phys. Rev. B {\bf 77}, 155420 (2008).

\bibitem{am07} M. P. Almeida, F. de Melo, M. Hor-Meyll, A. Salles, 
{\it et al.}, Science {\bf 316}, 579 (2007); A. Salles, F. Melo,
M. P. Almeida, M. Hor-Meyll, {\it et al.}, arXiv:0804.4556.

\bibitem{lc07} J. Laurat, K. S. Choi, H. Deng, C. W. Chou, and 
H. J. Kimble, Phys. Rev. Lett. {\bf 99}, 180504 (2007). 

\bibitem{brs1} S. Banerjee, V. Ravishankar and R. Srikanth, 
eprint arXiv:arXiv:0810.5034.

\bibitem{sqgen} R. Srikanth and S. Banerjee, 
Phys. Rev. A {\bf 77}, 012318 (2008); arXiv:0707.0059.

\bibitem{bd96} C. H. Bennett, D. P. DiVincenzo, J. A. Smolin and
W. K. Wootters, Phys. Rev. A {\bf 54}, 3824 (1996).

\bibitem{ww98} W. K. Wootters, Phys. Rev. Lett. {\bf 80}, 2245 (1998).

\bibitem{mc05} F. Mintert, A. R. R. Carvalho, M. Kus and A. Buchleitner, 
Phys. Reports {\bf 415}, 207 (2005).

\bibitem{fw89} R. F. Werner, Phys. Rev. A {\bf 40}, 4277 (1989).

\bibitem{br08} S. Bhardwaj and V. Ravishankar, 
Phys. Rev. A {\bf 77}, 022322 (2008).

\bibitem{chb96} C. H. Bennett, G. Brassard, S. Popescu, B. Schumacher
{\it  et al.}, Phys. Rev. Lett. {\bf 76}, 722 (1996).

\bibitem{bdcz98} H.-J.  Briegel, W. D\"ur,  J. I. Cirac and P. Zoller,
Phys.  Rev. Lett.  {\bf 81},  5932  (1998); W.  D\"ur, H.-J.  Briegel,
J. I. Cirac, and P. Zoller, Phys. Rev. A 59, 169 (1999).

\bibitem{ft02} Z. Ficek and R. Tana\'{s}, Phys. Rep. 
{\bf 372}, 369 (2002).

\bibitem{bsri06} S. Banerjee and R. Srikanth, 
to appear in Eur. Phys. J. D;  eprint quant-ph/0611161.

\bibitem{bsdiss} S. Banerjee and R. Srikanth,  Phys. Rev. A {\bf 76}, 
062109 (2007);  eprint: arXiv:0706.3633.

\bibitem{dicke} R. H. Dicke, Phys. Rev. {\bf 93}, 99 (1954). 

\bibitem{tbs02} T. E. Tilma, M. Byrd and E. C. G. Sudarshan, J. Phys. A:
Math. Gen. {\bf 35}, 9255 (2002).

\bibitem{dn04} C. M. Dawson and M. A. Nielsen, arXiv:quant-ph/0401061. 

\bibitem{lkpl} J. Lee, M. S. Kim, Y. J. Park and S. Lee,  
J. Mod. Optics {\bf 47}, 2151 (2000).

\bibitem{pr08} J. P. Paz and A. J. Roncaglia, arXiv:0809.1676; 
Phys. Rev. Lett. {\bf 100}, 220401 (2008).

\end{thebibliography}
\end{document}